\documentclass[twocolumn,prl,floatfix,superscriptaddress,nobibnotes]{revtex4-1}

\usepackage{amsmath}
\usepackage{amssymb}
\usepackage{graphicx}
\usepackage{hyperref}
\usepackage{color}

\begin{document}

\title{Evidence for spin-polarized bound states in \\ semiconductor-superconductor-ferromagnetic insulator islands}
\author{S.~Vaitiek\.{e}nas}
\affiliation{Center for Quantum Devices, Niels Bohr Institute, University of Copenhagen, 2100 Copenhagen, Denmark}
\author{R.~Seoane~Souto}
\affiliation{Center for Quantum Devices, Niels Bohr Institute, University of Copenhagen, 2100 Copenhagen, Denmark}
\affiliation{Division of Solid State Physics and NanoLund, Lund University, 22100 Lund, Sweden}
\author{Y.~Liu}
\affiliation{Center for Quantum Devices, Niels Bohr Institute, University of Copenhagen, 2100 Copenhagen, Denmark}
\author{P.~Krogstrup}
\affiliation{Center for Quantum Devices, Niels Bohr Institute, University of Copenhagen, 2100 Copenhagen, Denmark}
\author{K.~Flensberg}
\affiliation{Center for Quantum Devices, Niels Bohr Institute, University of Copenhagen, 2100 Copenhagen, Denmark}
\author{M.~Leijnse}
\affiliation{Center for Quantum Devices, Niels Bohr Institute, University of Copenhagen, 2100 Copenhagen, Denmark}
\affiliation{Division of Solid State Physics and NanoLund, Lund University, 22100 Lund, Sweden}
\author{C.~M.~Marcus}
\affiliation{Center for Quantum Devices, Niels Bohr Institute, University of Copenhagen, 2100 Copenhagen, Denmark}

\date{\today}

\begin{abstract}
We report Coulomb blockade transport studies of semiconducting InAs nanowires grown with epitaxial superconducting Al and ferromagnetic insulator EuS on overlapping facets. Comparing experiment to a theoretical model, we associate cotunneling features in even-odd bias spectra with spin-polarized Andreev levels. Results are consistent with zero-field spin splitting exceeding the induced superconducting gap. Energies of subgap states are tunable on either side of zero via electrostatic gates.
\end{abstract}

\maketitle


In hybrid quantum devices with both ferromagnetic and superconducting components, competition to align electron spins or pair them into singlets can result in complex ground states and corresponding electrical properties~\cite{Meservey1994, Buzdin2005, Sau2010, Eschrig2011, Li2014, Linder2015, Strambini2017, Bergeret2018, Manna2020}.
Recently, coexistence of proximity-induced superconductivity and ferromagnetism have been demonstrated in hybrid semiconducting nanowires~\cite{Vaitiekenas2020}.
Coulomb-blockade spectroscopy of superconducting quantum dots provides a window into subgap spectra~ \cite{Higginbotham2015} and their spin structure~\cite{Prada2020}.


Multiple Andreev scatterings at superconducting boundaries of a small normal conductor give rise to Andreev bound states (ABSs)~\cite{Prada2020}.
The states can carry supercurrent through the normal region and appear in tunneling spectroscopy as discrete levels below the superconducting gap \cite{Pillet2010, Nichele2020}.
Coulomb effects modify transport via ABSs~\cite{Grove2009, Chang2013}, for instance resulting in supercurrent reversal~\cite{vanDam2006, Jorgensen2007}.
When magnetic fields~\cite{Lee2014, Shen2018} or magnetic materials~\cite{Heinrich2018} are involved, spin-degenerate ABSs split and becomes spin selective, as seen in tunneling spectroscopy~\cite{Whiticar2021} and circuit quantum electrodynamics measurements~\cite{Hays2020}.
The spin-active interface between a superconductor and, for example, a ferromagnetic insulator~\cite{Tokuyasu1988} can also lead to spin-split ABSs~\cite{Hubler2012} or, in some cases, triplet superconductivity~\cite{Diesch2018}. 


\begin{figure}[b!]
\includegraphics[width=\linewidth]{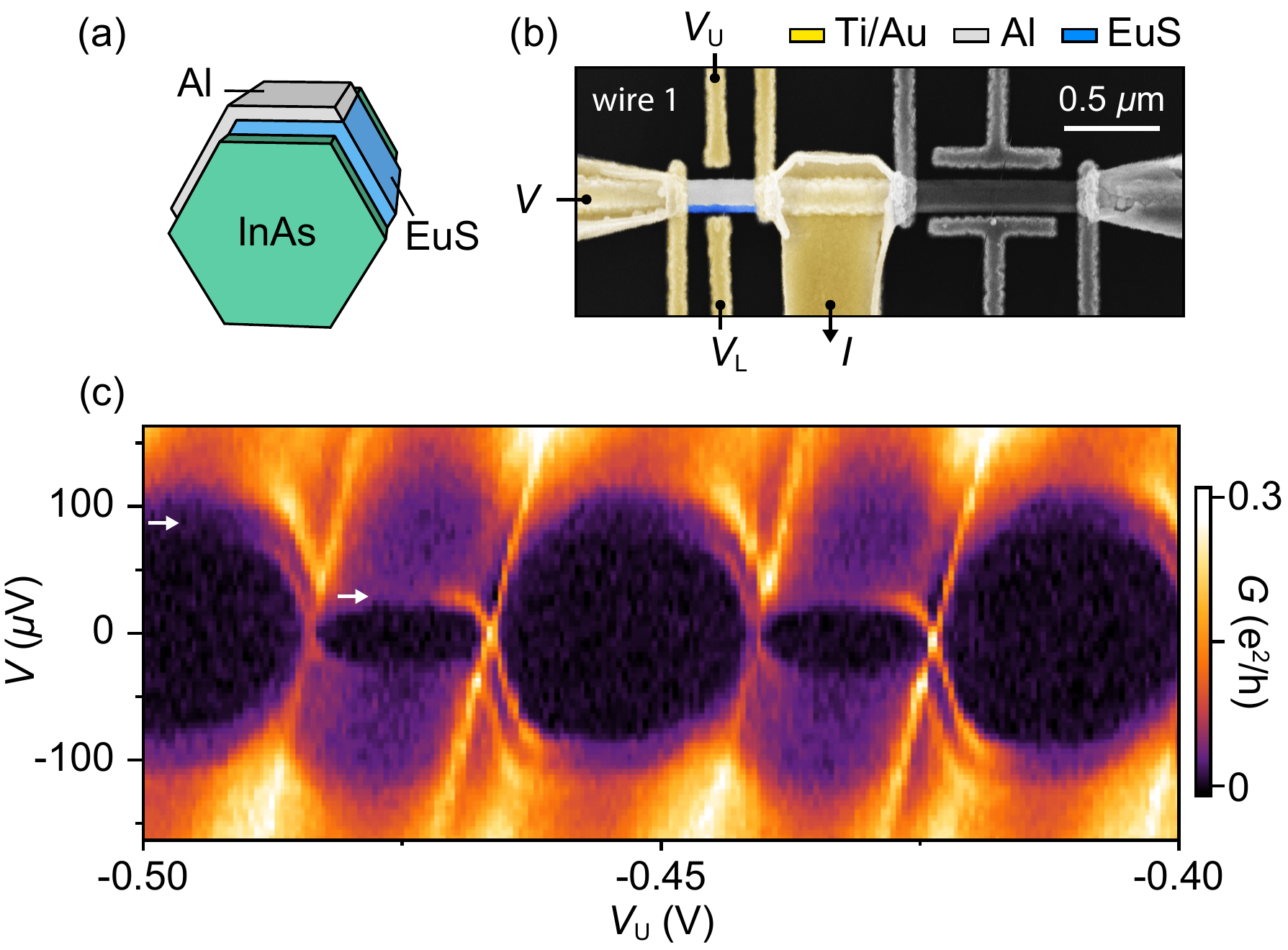}
\caption{\label{fig:1} 
(a) Schematic of hybrid InAs nanowire showing EuS and Al layers on overlapping facets of a hexagonal InAs nanowire.
(b) Scanning electron micrograph of wire~1. Enhanced false coloration highlights the 400~nm island measurement setup. 
(c) Differential conductance, $G$, as a function of voltage bias, $V$, and upper-gate voltage, $V_{\rm U}$, for the 400~nm island on wire 1 at zero applied magnetic field. Steps in conductance indicated by arrows alternate between $V = 30$ and $90~\mu$V.
}
\end{figure}

Recently, a new class of triple-hybrid materials was realized based on semiconducting InAs nanowires with strong spin-orbit coupling and large $g$ factor, coated with epitaxial superconducting Al, and ferromagnetic insulator EuS shells~\cite{Krogstrup2015, Liu2019_2}.
We investigate nanowires with hexagonal cross-sections and partly overlapping two-facet shells, as shown schematically in Fig.~\ref{fig:1}(a).
Tunneling spectroscopy into the ends of long grounded hybrid wires~\cite{Vaitiekenas2020} showed signatures consistent with topological superconductivity, as recently investigated theoretically~\cite{Woods2020, Maiani2020, Escribano2020, Liu2020, Langbehn2020, Khindanov2020}.

Here, we report transport through Coulomb islands, 400 and 800~nm in length, made from the same batch of wires with normal metal leads and several top- and side-gate electrodes that independently control tunnel-barrier conductances and charge occupancy~[Fig.~\ref{fig:1}(b)].
We observe characteristic features in Coulomb blockade that indicate gate-dependent, discrete subgap states whose energy can be tuned to zero. Qualitative comparison of cotunneling spectra to theoretical models suggest that the subgap states are spin polarized at zero magnetic field, as discussed in detail below.

Spectroscopy of four Coulomb island devices fabricated on two wires (denoted wire 1 and wire 2) showed similar results.
Measurements were carried out using standard low-noise lock-in techniques in a dilution refrigerator with a base temperature of 20~mK, equipped with a three-axis vector magnet (see Supplemental Material~\cite{Supplement}).

Differential conductance, $G=dI/dV$, of the 400~nm island on wire~1 as a function of source-drain voltage bias, $V$, and upper-gate voltage, $V_{\rm U}$, showed Coulomb diamonds of alternating height [Fig.~\ref{fig:1}(c)].
Once the tunneling barriers were coarsely tuned, this behavior is typical of all measured devices. 
Within a Coulomb valley, low-bias $G$ was suppressed below the experimental noise floor. At higher bias, $G$ showed a step-like increase at an alternating bias, as seen in Fig.~\ref{fig:1}(c).
The value of $V$ at which this first-step feature occurs could be tuned using the lower-gate voltage, $V_{\rm L}$.
A less pronounced second step in $G$ at higher bias [around $V = 120~\mu$V in Fig.~\ref{fig:1}(c)] did not alternate from valley to valley nor varied with $V_{\rm L}$ (see Fig.~S1 in Supplemental Material~\cite{Supplement}).
The charging energy, $E_{\rm C} = 300~\mu$eV, measured from the Coulomb diamonds, is larger than the superconducting gap of the parent Al shell, $\Delta_{\rm Al} = 230~\mu$eV hence also larger than the induced gap, $\Delta$, which is reduced by the coupling to EuS~\cite{Vaitiekenas2020}.
The 800~nm island on wire 1 showed similar even-odd periodic Coulomb blockade with step-like cotunneling features at finite bias  (see Fig.~S2 in Supplemental Material~\cite{Supplement}).


To understand the conductance features and relate them to ABSs and spin, we model transport through a superconducting Coulomb island, including a single subgap state, spin-split by Zeeman energy, $E_{\rm Z}$.
Sequential single-electron tunneling through an ABS on the island yields characteristic Coulomb diamonds~\cite{Higginbotham2015, VanHeck2016}.
To account for intermediate strengths of tunnel couplings to both leads we also include cotunneling processes~\cite{Ekstrom2020} through a next-to-leading order expansion in the T-matrix (see Supplemental Material~\cite{Supplement}).
Elastic cotunneling gives a bias independent background conductance, while inelastic cotunneling leaves the system in an excited state yielding steps in $G$ when the bias matches excitation energies.

\begin{figure}[t!]
\includegraphics[width=\linewidth]{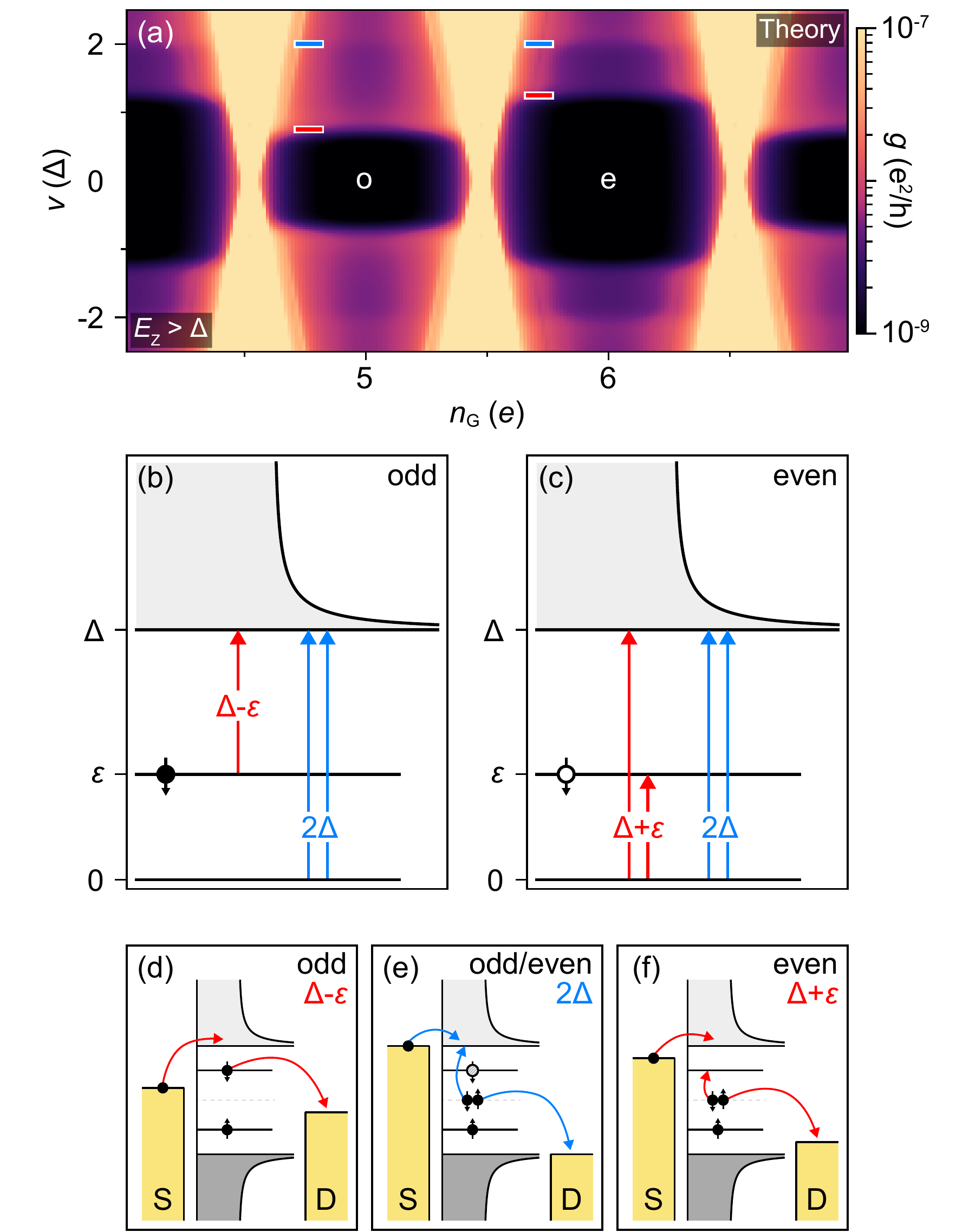}
\caption{\label{fig:2} 
(a) Calculated conductance, $g$, as a function of voltage bias, $v$, and charge offset, $n_{\rm G}$, for a superconducting island with a superconducting gap $\Delta$, charging energy $E_{\rm C} = 5 \Delta$, Zeeman energy $E_{\rm Z}\gg\Delta$ (shown here for $E_{\rm Z} = 100\Delta$), and a spin-polarized subgap state at energy $\varepsilon = \Delta/4$ (see Supplemental Material~\cite{Supplement}).
Steps in $g$ due to cotunneling are visible at $v=\Delta-\varepsilon$ in odd (o) and $v=\Delta+\varepsilon$ in even (e) valleys as well as at $v=2\Delta$ in both valleys.
The calculations were done for finite temperature $k_{\rm B}\,T=0.03\,\Delta$.
(b) and (c) Schematic superconducting density of states for odd (b) and even (c) valleys in (a) with excitations indicated by arrows.
(d) In the odd ground state, a quasiparticle in the bound state exits the island, while another quasiparticle from a lead tunnels into the continuum. The excitation energy of $\Delta-\varepsilon$ is supplied by the voltage bias.
(e) In both valleys, a voltage bias of $2\Delta$ can break a Cooper pair. In that case, one quasiparticle leaves the island, while the other, together with a quasiparticle from a lead, is excited to the continuum.
(f) In the even ground state, a spin-down quasiparticle from a broken Cooper pair is excited into the empty bound state and the spin-up quasiparticle leaves the island, while another quasiparticle tunnels into the continuum. The total energy cost of the process is $\Delta+\varepsilon$.  
}
\end{figure}

Theoretical values for differential conductance, $g$, of a Coulomb island containing a single spin-split ABS as a function of voltage bias, $v$, and gate-induced charge offset, $n_{\rm G}$, is shown in Fig.~\ref{fig:2}(a), where $\varepsilon_\uparrow>\Delta$ and $\varepsilon_\downarrow\equiv\varepsilon$ are energies of the two spin branches.  
The main experimental features are captured qualitatively in this simple theoretical model. In particular, the bias value of the conductance step alternates between even (e) and odd (o) island parities. Steps in differential conductance, marked by the red and blue ticks, correspond to transitions between ground and lowest excited states, as illustrated in Figs.~\ref{fig:2}(b) and \ref{fig:2}(c).
Red lines marking $\Delta - \varepsilon$ for odd valleys and $\Delta + \varepsilon$ for even valleys correspond to processes that change the parity of the subgap state, while the blue lines  marking $2\Delta$ for both valleys correspond to processes that break Cooper pairs without changing parity.
Cotunneling processes involving higher-energy intermediate states with $\pm1$ charge on the island are shown in Figs.~\ref{fig:2}(d)--\ref{fig:2}(f).

Theoretical spectra for spin-degenerate or weakly spin-split ABS show a denser pattern of cotunneling steps associated with excitations to spin-flipped states at fixed charge, as shown in Fig.~S3 in Supplemental Material~\cite{Supplement}. 
A qualitative comparison shows that the experimental data from Fig.~\ref{fig:1}(c) agree better with the spin-polarized rather than the spin-degenerate theory (see the discussion and Figs.~S4 and S6 in Supplemental Material~\cite{Supplement}).
Previous measurements of a similar hybrid island, but without the EuS shell, had shown Coulomb spectra that are consistent with the theory for a spin-degenerate bound state, see Fig.~S7 in Supplemental Material~\cite{Supplement}.


Returning to experiment, transport data for the 400~nm device on wire 2 shown in Fig.~\ref{fig:1}(c) yield $\Delta+\varepsilon = 120~\mu$eV and $\Delta-\varepsilon = 60~\mu$eV, giving $\varepsilon = 30~\mu$eV and $\Delta = 90~\mu$eV, consitent with the $2\Delta$ feature at $180~\mu$eV.
The data in Fig.~\ref{fig:1}(c) gives a slightly smaller $\Delta = 60~\mu$eV. We note that $\Delta$ can be gate-voltage dependent. In general, the induced gap is considerably smaller than the parent Al gap in these wires. The deduced $E_{\rm C} = 430~\mu$eV is larger than $\Delta$, consistent with the even-odd periodic Coulomb pattern. The sharp spectral features at the degeneracy points indicate a discrete subgap state.

\begin{figure}[t!]
\includegraphics[width=\linewidth]{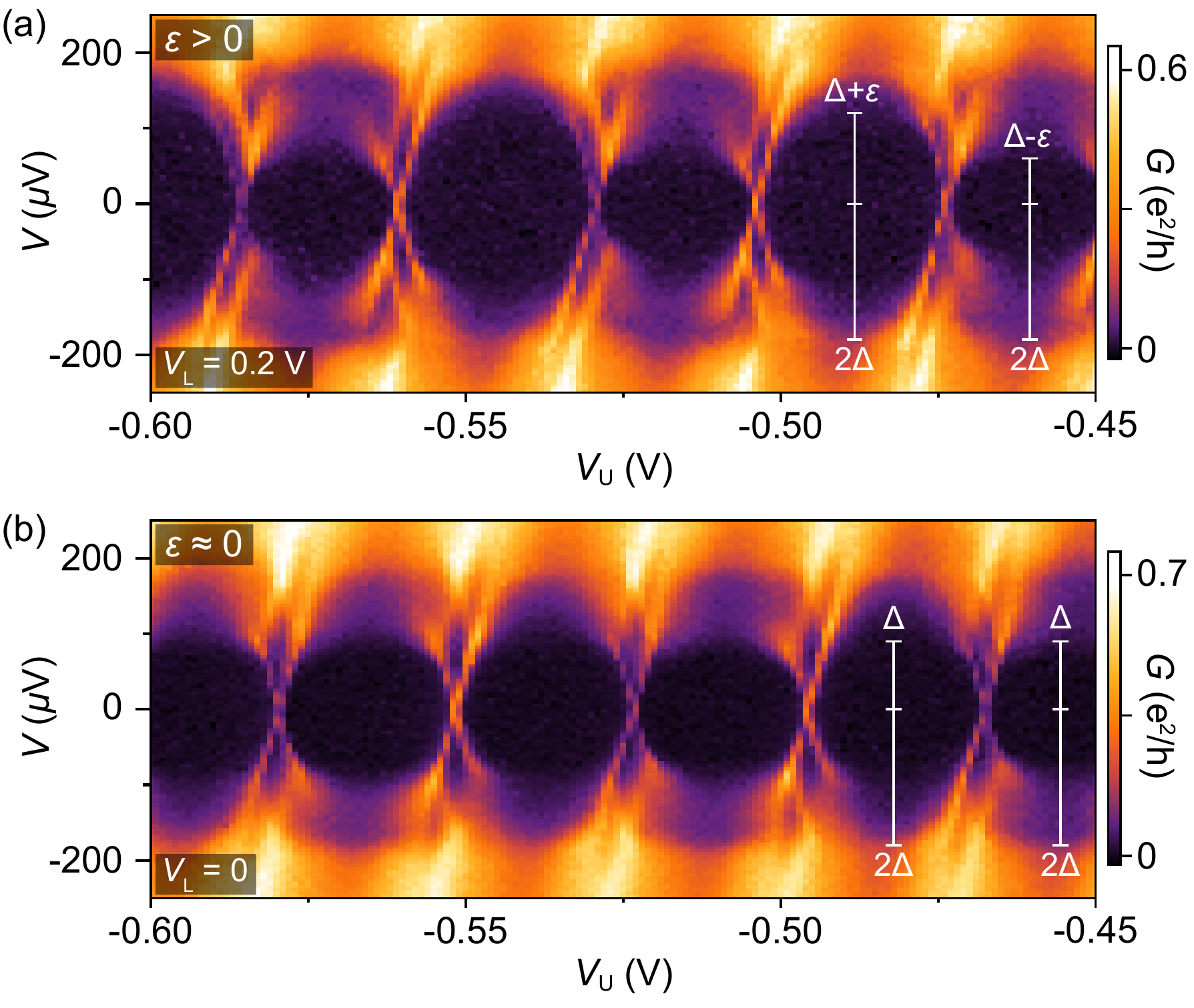}
\caption{\label{fig:3} 
(a) Differential conductance, $G$, as a function of source-drain bias, $V$, and upper gate voltage,\ $V_{\rm U}$, for the 400 nm island on wire 2.
A clear even-odd Coulomb diamond pattern is visible with an inelastic onset in $V$ at $\Delta + \varepsilon = 120~\mu$eV for the bigger diamond and $\Delta - \varepsilon = 60~\mu$eV for the smaller one, as well as an additional step at $2\Delta = 180~\mu$eV in both diamonds, giving $\Delta = 90~\mu$eV and $\varepsilon = 30~\mu$eV.
The data were taken at a fixed lower-gate voltage $V_{\rm L} = 0.2$~V.
(b) Similar to (a) measured at $V_{\rm L} = 0$, giving nearly 1$e$-periodic Coulomb diamonds with two steps in $G$ for all diamonds at $\Delta = 90~\mu$eV and $2\Delta = 180~\mu$eV, indicating $\varepsilon \approx 0$.
The measured charging energy $E_{\rm C} = 430~\mu$eV.
}
\end{figure}

Decreasing $V_{\rm L}$ from +0.2~V to 0 modifies the Coulomb blockade peaks from distinctly even-odd to $1e$-periodic at zero bias, with consecutive diamonds differing only by the intensity of step features at finite bias, as seen in Fig.~\ref{fig:3}(b).
The onsets of the lower-energy steps in both valleys align with $\Delta = 90~\mu$eV, indicating that $\varepsilon \approx 0$.

We interpret the evolution as reflecting a subgap state that gradually decreases to zero energy as $V_{\rm L}$ is varied.
In other words, the gate-induced electric fields change the electrostatic environment of the hybrid nanowire, thus modifying the parameters of the subgap state and effectively changing its energy.
In the present context, the evolving spin-mixing angle at the superconductor-ferromagnetic insulator interface~\cite{Hubler2012} contributes to the gate dependence of $\varepsilon$.

Similar measurements at various $V_{\rm L}$ are shown in Fig.~S8 in Supplemental Material~\cite{Supplement}.
The even-odd structure in the amplitude of the finite bias conductance steps is expected theoretically, and reflects the relative phase difference between electron and hole components of the subgap state (see the discussion and Fig.~S9 in Supplemental Material~\cite{Supplement}).

\begin{figure*}[t!]
\includegraphics[width=\linewidth]{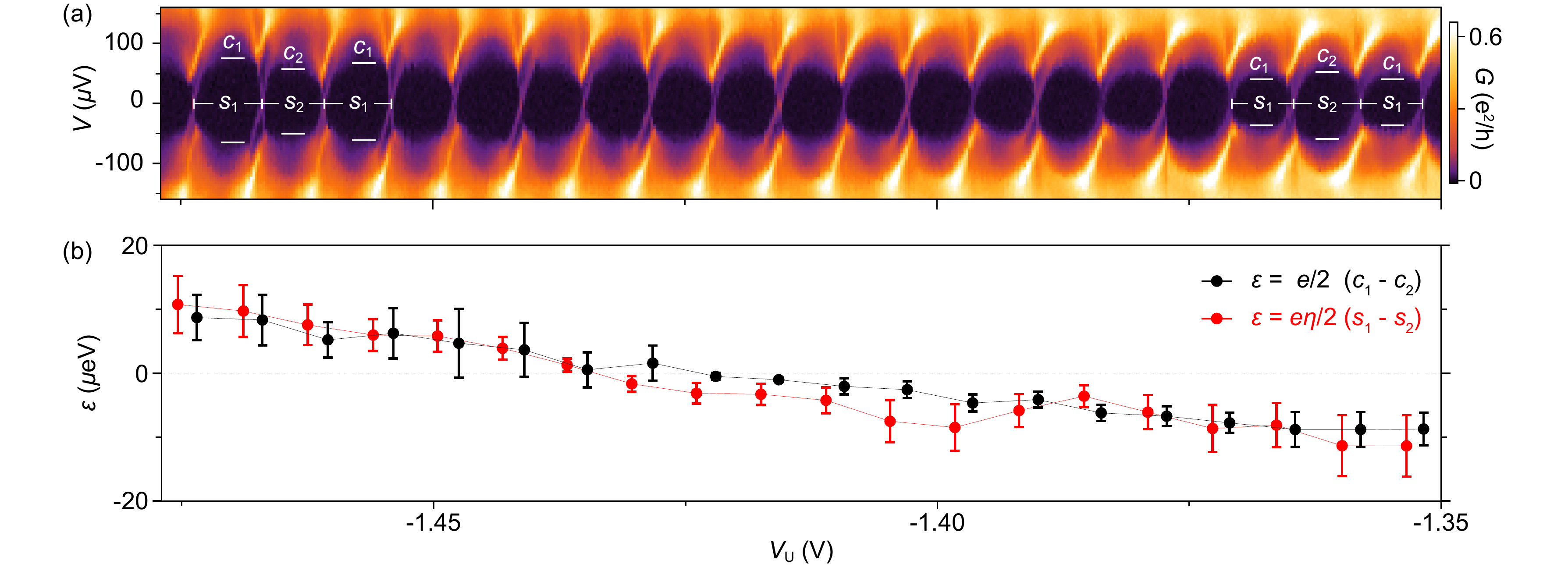}
\caption{\label{fig:4} 
(a) Differential conductance, $G$, measured for the 800 nm island on wire 2 as a function of source-drain bias, $V$, over an extended range of the upper-gate voltage, $V_{\rm U}$. The Coulomb blockade pattern evolves from even-odd at $V_{\rm U} = -1.47$~V, through 1$e$ around $V_{\rm U} = -1.43$~V, to even-odd periodicity again at $V_{\rm U} = -1.35$~V, visible in both inelastic cotunneling onsets $c_{\rm i}$ and peak spacings $s_{\rm i}$, where $i = 1$ and 2 denote Coulomb valleys.
The larger $i=1$ diamonds on the negative side of the measured gate voltage range become smaller than the $i=2$ diamonds on the more positive side.
The measured charging energy $E_{\rm C} = 320~\mu$eV is smaller than for the 400~nm island.
The data were taken at $V_{\rm L} = -0.7$~V.
(b) Subgap state energy, $\varepsilon$, inferred from step heights (black) and peak-spacing differences (red).
$\varepsilon$ decreases monotonically from roughly $10~\mu$eV through 0 to $-10~\mu$eV as the gate voltage is increased.
Black error bars represent standard errors from the $c_{\rm i}$ measurement at the positive and negative $V$; red error bars were estimated by propagation of error from the Lorentzian peak fitting and lever arm, $\eta$, measurement.
}
\end{figure*}

To investigate $\varepsilon$ dependence on the electrostatic environment we look over a wider range of gate voltages. The observed even-odd pattern crosses smoothly through 1$e$ periodicity, reflecting the continuous evolution of $\varepsilon$ across zero.
This is shown in Fig.~\ref{fig:4}(a) as a function of upper-gate voltage, $V_{\rm U}$, for the 800~nm island on wire~2.
Both the onsets of the high-bias features, at values $c_{\rm i}$, and the peak spacings, $s_{\rm i}$, alternate in magnitude.
Subscripts $i=1$ and 2 denote the two different charge occupancies of the island.
We define $c_1 = (\Delta + \varepsilon)/e$ and $c_2 = (\Delta - \varepsilon)/e$, then take the difference between consecutive $c_{\rm i}$ to extract the subgap-state energy $\varepsilon$, as a function of $V_{\rm U}$ as shown in Fig.~\ref{fig:4}(b).
Within the measured range of $V_{\rm U}$, $\varepsilon$ decreases monotonically from $+10~\mu$eV to $-10~\mu$eV.
Independently, values for $\varepsilon$ were extracted from Coulomb peak spacing at zero bias~\cite{Higginbotham2015, Albrecht2016, Shen2018}. For $E_{\rm C} > \Delta > \varepsilon$, peak spacings are given by $s_1 = (E_{\rm C} + \varepsilon)/e\eta$ and $s_2 = (E_{\rm C} - \varepsilon)/e\eta$, where $\eta$ is a dimensionless lever arm measured from the slopes of the Coulomb diamonds.
The subgap-state energy inferred from the Coulomb peak spacing difference agrees well with $\varepsilon$ deduced from finite bias steps, as shown in Fig.~\ref{fig:4}(b).
Good qualitative agreement between measured and computed spectra is shown in Fig.~S10 in Supplemental Material~\cite{Supplement}.
Similar analysis for $\varepsilon$ as a function of $V_{\rm L}$, where the subgap state approaches but does not cross zero energy, is shown in Fig.~S11 in Supplemental Material~\cite{Supplement}.

The sign of $\varepsilon$ depends on the definition of $c_i$ and $s_i$.
Assuming strong zero-field spin splitting at zero applied magnetic field leaves it ambiguous whether or not a level has crossed zero energy. We therefore cannot label the even and odd valleys with certainty.
We note that while in principle the evolution of $\varepsilon$ with applied magnetic field contains information on the spin projection of the bound state and hence the ground state parity, we are not able to determine if the field predominantly affects the Zeeman splitting or the magnetization of the EuS (see Ref.~\cite{Vaitiekenas2020}). Representative magnetic-field data for both islands on wire 2 are shown in Figs.~S11 and S12 in Supplemental Material~\cite{Supplement}.

Finally, we note that for specific gate configurations there are no inelastic cotunneling steps present in Coulomb diamonds (see Fig.~S14 in Supplemental Material~\cite{Supplement}).
This can be understood within the model as resulting from the condition $\varepsilon > \Delta$, yielding a cotunneling background for all voltage-bias values within odd valleys and non-zero conductance above $2\Delta$ in even valleys.

We thank Claus S\o rensen for contributions to materials growth and Shivendra Upadhyay for assistance with nanofabrication. We acknowledge support from the Danish National Research Foundation, European Research Council (Grants Agreement No. 716655 and No. 856526), Microsoft, NanoLund, QuantERA (Project "2D hybrid materials as a platform for topological quantum computing"), and a research grant (Project 43951) from VILLUM FONDEN. 

\onecolumngrid
\twocolumngrid
\bibliography{bibfile}

\onecolumngrid
\clearpage
\onecolumngrid
\setcounter{figure}{0}
\setcounter{equation}{0}
\section{Supplemental Material}
\renewcommand\thefigure{S\arabic{figure}}
\renewcommand{\tablename}{Table.~S}
\renewcommand{\thetable}{\arabic{table}}
\twocolumngrid

\section{Sample preparation}
InAs wires with hexagonal cross section were grown to a length of $\sim 10~\mu$m and diameter of $\sim 120$~nm using molecular beam epitaxy~\cite{Krogstrup2015}.
Partly-overlapping, two-faceted EuS (as grown thickness 8~nm) and Al (as grown thickness 6~nm) shells were grown \textit{in~situ} using electron beam evaporation, as shown in the main-text Fig.~1(a)~\cite{Liu2019,Liu2019_2}.
Devices were fabricated on a Si substrate with 200~nm SiOx capping.
Coulomb islands were formed by wet-etching (MF-321 photoresist developer, 30~s, room temperature) and contacting the exposed InAs/EuS with Ti/Au (5/150~nm) after \textit{in~situ} Ar milling (15~W, 7~min).
Devices were then coated with a HfO$_2$ (8~nm) dielectric layer, followed by the deposition of Ti/Au (5/150~nm) gates patterned using electron beam lithography. Additional details can be found in Ref.~\cite{Vaitiekenas2020}.

\section{Measurements}
The samples were cooled to base temperature of the dilution refrigerator at zero applied magnetic field.
Following cooldown, an external magnetic field was applied $\mu_0 H_\parallel = 150$~mT along the wire axis then returned to zero.
Unless otherwise noted, the measurements were carried out at zero applied field.
Differential conductance, $G=$~d$I/$d$V$, was measured by sourcing voltage bias, $V$, through one of the outer leads, floating the opposite end of the wire, and draining the current, $I$, through the common lead between the two islands, as shown for the 400~nm island in the main-text Fig.~1(b) and for the 800~nm island in Fig.~\ref{fig:S2}(a).
The islands were tuned into the Coulomb blockade regime using negative voltages applied to the cutter gates, $V_{\rm C1}$ and $V_{\rm C2}$, and back gate, $V_{\rm BG}$ [labeled only in Fig.~\ref{fig:S2}(a)].
Over a range much larger than the Coulomb peak spacing, $V_{\rm BG}$ also tuned the chemical potential of the island.
The upper-gate voltage, $V_{\rm U}$, on the side coated with the Al shell, was predominantly used to tune the occupancy of the hybrid island, whereas the lower-gate voltage, $V_{\rm L}$, on the EuS shell side, was used to tune the charge carrier density in the semiconductor.

\section{Model}
We consider a superconducting island hosting a subgap state that can be spin-slit (for example, due to exchange-coupling to the ferromagnetic insulator) by an energy $E_{\rm Z}$.
For simplicity, we describe the continuum of states as a spin-degenerate quasiparticle state at energy $\Delta$, using the so-called zero bandwidth model~\cite{Grove2018}.
We take the charging energy of the island, $E_{\rm C}$, to be the largest and the coupling to the leads to be the smallest energy scales in the system, allowing us to treat the electron tunneling in a perturbative way.

The Hamiltonian of the system is given by
    \begin{equation}
    	H=H_{\rm L}+H_{\rm I}+H_{\rm T}\,,
    \end{equation}
where the leads are described by 
    \begin{equation}
	    H_{\rm L}=\sum_{\rm \nu,k,\sigma}\xi_{\nu k\sigma}c^{\dagger}_{\nu k\sigma}c_{\nu k\sigma}\,,
    \end{equation}
with the energy, $\xi_{\nu k\sigma}$, and the annihilation operator, $c_{\nu k\sigma}$, of an electron in lead $\nu \in \{L, R\}$ with momentum $k$ and spin $\sigma \in \{\uparrow, \downarrow\}$. We assume that each lead remains in internal equilibrium described by a Fermi-Dirac distribution with chemical potential $\mu_\nu$. 

The superconducting island is described by
    \begin{equation}
    	H_{\rm I}=\sum_{j,\sigma}\vartheta_{j\sigma} \gamma_{j\sigma}^\dagger \gamma_{j\sigma}+E_{\rm el}(N)\,,
    	\label{H_island}
    \end{equation}
where $j \in \{0,\Delta\}$ is the state index, $\vartheta_{j\sigma}$ is the (spin-dependent) energy of subgap and continuum states, $\gamma_{j\sigma}$ is the Bogoliubov quasiparticle annihilation operator, and $E_{\rm el}$ is the electrostatic repulsion term that depends on the number of electrons in the island, $N=2N_{\rm C}+n_{\rm qp}$, where $N_{\rm C}$ is the number of Cooper pairs and $n_{\rm qp}$ is the total charge in the quasiparticle states.
For the subgap state, we take $\vartheta_{0\downarrow}=\varepsilon$ and $\vartheta_{0\uparrow}=\varepsilon+E_{\rm Z}$.
For simplicity, we take spin-degenerate continuum states at $\vartheta_{1\downarrow} = \vartheta_{1\uparrow} = \Delta$.
Similar results are found for spin-split continuum states.
The annihilation operator for an electron in the island, $d_{j\sigma}$, is related to $\gamma_{j\sigma}$ by
    \begin{equation}
        \gamma_{j\sigma}=u_j d_{j\sigma}-\rho_\sigma \, v_je^{-i\phi}d_{j\bar{\sigma}}^\dagger\,,
    \end{equation}
where $u_j$ and $v_j$ are the Bogoliubov coefficients, $\rho_{\uparrow/\downarrow}=\pm1$ depending on the spin, $\phi$ is the superconducting phase operator such that $e^{i\phi}$ creates a Cooper pair on the island, and $\bar{\sigma}$ is the spin opposite to $\sigma$.
The electrostatic repulsion term is given by
    \begin{equation}
    	E_{\rm el}(N)=E_{\rm C}(N-n_{\rm G})^2\,,
    \end{equation}
where $n_{\rm G}$ is the dimensionless gate-induced charge offset.

The tunneling Hamiltonian is described by
    \begin{equation}
        \begin{split}
    	H_T & =\sum_{\nu, k,j,\sigma} \left(t_{\nu k\sigma}c_{\nu k\sigma}^\dagger d_{j\sigma}+h.c.\right)\\
    	& =\sum_{\nu, k,j,\sigma} \left[t_{\nu k\sigma}c_{\nu k\sigma}^\dagger \left(u_j\gamma_{j\sigma}+\sigma v_j e^{-i\phi}\gamma_{j\bar{\sigma}}^\dagger\right)+h.c.\right]\,,
    	\end{split}
    \end{equation}
where $t$ denotes the tunneling amplitudes between the island and leads.

\section{Formalism}
The transport properties through the superconducting island are calculated using the T-matrix formalism \cite{Bruus2004}.
The transition rate between initial, $\left|i\right\rangle$, and final, $\left|f\right\rangle$, states can be computed by
    \begin{equation}
    	\Gamma_{i\to f}=2\pi\left|\left\langle f\left| T \right|i\right\rangle \right|^2 W_{if}\delta(E_i-E_f)\,,
    	\label{rates_def}
    \end{equation}
where $W_{if}$ weights the rate through thermal distributions of initial and final states at energies $E_i$ and $E_f$, respectively, $\delta$ is the Dirac delta function and
    \begin{equation}
    	T=H_{\rm T}+H_{\rm T}\frac{1}{E_i-H_{\rm L}-H_{\rm I}+i0^+}T\,,
    	\label{T_definition}
    \end{equation}
which can be truncated at the desired order.
Here the term linear in $H_{\rm T}$ describes the sequential tunneling; the higher order terms describe the cotunneling contributions, which become progressively more important when the tunneling coupling between the island and the leads increases.

The quantum state of the island is described by $\left|a\right\rangle$=$\left|N,N_{\rm C},\textbf{n}\right\rangle$, where $\textbf{n}$ is a vector representing the occupation of the subgap and continuum excited states. The time derivative of the occupation probability of a given state can be written as
    \begin{equation}
    	\dot{P}_a=\sum_b\left[-\Gamma_{a\to b}P_a+\Gamma_{b\to a}P_b\right]\,,
    \end{equation}
which describes the stationary probability of occupation, $P^{\rm{stat}}_b$, by imposing $\dot{P}^{\rm{stat}}_b=0$ and $\sum_b P^{\rm{stat}}_b=1$.
The current through the device can be computed using the stationary distribution of probabilities and the rates.
The tunneling current is given by
    \begin{eqnarray}
        \begin{split}
    	 I=\sum_{a,b}\left[\frac{s}{2}\Gamma^{\rm{seq}}_{b\to a}
	     +\delta n(L\to R)\,\Gamma^{\rm{cot}}_{b\to a} \vphantom{\frac{1}{2}}\right]P^{\rm{stat}}_b\,,
        \end{split}
    \end{eqnarray}
where $s=+1$ for rightwards sequential tunneling and $-1$ for the opposite direction, $\delta n(L\to R)$ is the net charge transferred from left to right in a cotunneling process, and the sum runs over all the possible rates connecting the island state $\left|b\right\rangle$ with any state $\left|a\right\rangle$.

\section{Sequential tunneling rates}
The sequential rates are given by
    \begin{align}
    \begin{split}
         	\Gamma^{\rm{seq}}_{\left|N,N_{\rm C},\textbf{n}\right\rangle\to\left|N+1,N_{\rm C},\textbf{n'}\right\rangle}&=\Gamma_\nu \left|u_j\right|^2 n_{\rm F}(E_f-E_i-\mu_\nu)\,\\
	        \Gamma^{\rm{seq}}_{\left|N,N_{\rm C},\textbf{n}\right\rangle\to\left|N-1,N_{\rm C},\textbf{n'}\right\rangle}&=\Gamma_\nu \left|u_j\right|^2 n_{\rm F}(\mu_\nu+E_f-E_i)\,\\
	        \Gamma^{\rm{seq}}_{\left|N,N_{\rm C},\textbf{n}\right\rangle\to\left|N+1,N_{\rm C}+1,\textbf{n'}\right\rangle}&=\Gamma_\nu \left|v_j\right|^2 n_{\rm F}(E_f-E_i-\mu_\nu)\,\\
         	\Gamma^{\rm{seq}}_{\left|N,N_{\rm C},\textbf{n}\right\rangle\to\left|N-1,N_{\rm C}-1,\textbf{n'}\right\rangle}&=\Gamma_\nu \left|v_j\right|^2 n_{\rm F}(\mu_\nu+E_f-E_i)\,,
    \end{split}
    \end{align}
where $n_{\rm F}$ is the Fermi-Dirac distribution function and $\mu_\nu$ is the chemical potential of the lead $\nu$.
Here, we have used the wideband approximation, where the tunneling rates are energy independent and $\Gamma_\nu=2\pi\rho_F |t_\nu|^2$, with the lead density of states at the Fermi level $\rho_F$.
Note that for these rates $\textbf{n}$ and $\textbf{n}'$ differ by the occupation of one state.

\section{Cotunneling rates}
We consider the cotunneling rates transferring an electron from one lead to the other, which are the dominant ones in the limit $E_{\rm C}\gg\max\left(\Delta,\,v \equiv \mu_{\rm L} - \mu_{\rm R}\right)$, where $v$ is the source-drain voltage bias used in calculations.
These processes, which conserve the charge on the island, can be expressed by
    \begin{equation}
        \begin{split}
    	F(\omega_1) & \equiv \left|\left\langle f\left| T \right|i\right\rangle\right|^2 = \\
    	& \left| \sum_{m_1,\nu}\frac{\left\langle f\left| H_{\rm T}(\nu) \right|m_1\right\rangle\left\langle m_1\left| H_{\rm T}(\bar{\nu}) \right|i\right\rangle}{E_{m_1}-E_i-\omega_1} \right. \\
    	&\left. +\sum_{m_2,\nu} \frac{\left\langle f\left| H_{\rm T}(\bar{\nu}) \right|m_2\right\rangle\left\langle m_2\left| H_{\rm T}(\nu) \right|i\right\rangle}{E_{m_2}-E_f+\omega_1}\right|^2\,,
        \end{split}
        \label{Kernel_2order}
    \end{equation}
where $H_{\rm T}(\nu)$ describes the tunneling between $\nu$ lead and the island, $\bar{\nu}$ denotes a lead opposite to $\nu$, and  $m_{1,2}$ are virtual intermediate states. To derive this expression we have imposed energy conservation, which leads to a function dependent on the energy of the tunneling electron from/to one of the leads, $\omega_{1}$.
The cotunneling rate can be written as
    \begin{equation}
        \begin{split}
        	\Gamma^{\rm{cot}}_{i\to f}=2\pi\int d\omega_1\,& F(\omega_1)\, n_{\rm F}(\omega_1-\mu_{\rm L})\,\\
        	\times\, & n_{\rm F}(\mu_{\rm R}+E_f-E_i-\omega_1)\,.
        	\label{cot_rate}
        \end{split}
    \end{equation}
This expression for the cotunneling rates is divergent due to the appearing sequential tunneling. To avoid the divergent behaviour, we regularize the divergences as explained in Ref. \cite{Koller2010}. The resulting integral can be formally solved analytically~\cite{Koch2004}, which leads to a complicated expression involving special functions. In the limit of $T/E_{\rm C}\gtrsim10^{-3}$, it turns out to be  more computationally efficient to simplify it by expanding $n_{\rm F}$ into a sum of complex Matsubara-Ozaki frequencies~\cite{Ozaki2007}
\begin{equation}
	n_{\rm F}(\omega-\mu)-\frac{1}{2}=\sum_\alpha r_\alpha\frac{1}{\omega-\mu+i\beta_\alpha}\,,
\end{equation}
where $\beta_\alpha$ and $r_\alpha$ are the approximated Matsubara frequencies and residues, respectively.
Finally, Eq.~\eqref{cot_rate} can be evaluated using the residue theorem yielding
    \begin{equation}
        \begin{split}
    	\Gamma^{\rm{cot}}_{i\to f} = & \,8\pi\, \mbox{Im}\sum_{\alpha=0}^\infty r_\alpha \left\{ n_{\rm F}(\mu_{\rm R}+E_f-E_i-\mu_{\rm L}+i\beta_\alpha) \right. \\ 
    	 &\left. \times \left[F(\mu_{\rm L}-i\beta_\alpha) - F(\mu_{\rm R}+E_f-E_i+i\beta_\alpha)\right] \right\}\,.
        \end{split}
    \end{equation}
This sum can be truncated at $\alpha\approx 100$ Matsubara frequencies for the parameters used in the calculations.

\section{Transport calculations}
Calculated Coulomb spectra for a spin-degenerate ($E_{\rm Z} = 0$) and weakly spin-split ($E_{\rm Z} < \Delta$) subgap states are shown in Figs.~\ref{fig:S3}(a) and \ref{fig:S3}(g).
The colored ticks mark the onset of inelastic cotunneling events that excite the system into a higher-energy state, resulting in a step-like increase in conductance.
The transitions in the odd and even valleys are represented separately in panels (b) and (c) for the spin-degenerate, and panels (h) and (i) for the weakly spin-split cases.
Some examples of the cotunneling transport mechanisms for the two cases are illustrated in panels (d)--(f) and (j)--(l), respectively.
In general, there are four cotunneling lines in both spin-degenerate and weakly spin-split cases (except for the specific instances where $\varepsilon$ is fine tuned such that two transitions are degenerate in energy).
In contrast, our experimental data show only two cotunneling steps (see, for example, Fig.~3 in the main text). A qualitative line-cut comparison between experimental data and spin-polarized as well as spin-degenerate cases is shown for the 400 nm island on wire 1 in Fig.~\ref{fig:S4}.
Aggregated line-cuts from several consecutive even and odd Coulomb valleys measured for the 400 nm island on wire 2 display two cotunneling steps, independent of lower-gate voltage and subgap-state energy, see Fig.~\ref{fig:S5}.

The measured cotunneling features decrease in energy without splitting as an external magnetic field is applied, see Fig.~\ref{fig:S6}(b). For comparison, we calculate conductance for a spin-polarized and two spin-degenerate cases---with and without magnetic hysteresis. As an input, we use approximated expressions for the field dependence of the superconducting gap and the subgap states. We describe the gap closing as $\Delta(H_\parallel)=\Delta_0\left[1-\left(H_\parallel/H_{\rm C}\right)^2\right]$, with critical field $H_{\rm C}=60$~mT and the zero-field gap $\Delta_0=60~\mu$eV, see the black curves in Fig.~\ref{fig:S6}(c), (e), and (g). For simplicity we assume that the subgap state depends linearly on the field, see the red and blue curves in Fig.~\ref{fig:S6}(c), (e), and (g). To account for the asymmetry in the measured data, we include magnetic hysteresis of 10~mT (consistent with previous experiments~\cite{Vaitiekenas2020}) in the spin-polarized and one of the spin-degenerate cases. The magnetic field in the spin-degenerate case leads to the appearance of additional cotunneling features that are not observed experimentally, see Fig.~\ref{fig:S6}(f) and (h). We therefore find that the measurements agree best with the spin-polarized spectrum shown in Fig.~\ref{fig:S6}(d).

We note that the number of cotunneling steps increases in similar hybrid island devices without EuS shell (see the replotted data from Ref.~\cite{Higginbotham2015} in Fig.~\ref{fig:S7}) consistent with Fig.~\ref{fig:S3}(a).

These observations together suggest that the investigated hybrid islands are in the strongly spin-split limit ($E_Z\gtrsim\Delta$), discussed in the main-text Fig.~2.

The relative height of the cotunneling steps in the even and odd Coulomb valleys depends on the relative phase, $\varphi$, between the Bogoliubov components of the subgap state, $u_0 \equiv \vert u_0 \vert\exp(i\varphi)$ and $v_0 \equiv \vert v_0 \vert$, as illustrated in Fig.~\ref{fig:S9} for $E_{\rm Z}\gg\Delta$ and $\varepsilon = 0$.
The effect can be explained by the interference between different cotunneling mechanisms that involve the same initial and final states.
However, $\varphi$ cannot be determined unambiguously in our setup, as the strong spin splitting and gate-dependent $\varepsilon$ makes the global ground state unknown.
This remains an open problem, relevant for distinguishing trivial and topological states.

In the experiment, the magnitude and, in some cases, the sign of $\varepsilon$ can be tuned using electrostatic gate electrodes (Fig.~4 in the main text).
In our model, we account for this behavior by changing $\vartheta_{0 \sigma}$ in Eq.~\eqref{H_island}.
A change in the sign of $\varepsilon$ is equivalent to exchanging the Bogoliubov components of the corresponding subgap state.
In Fig.~\ref{fig:S10} we show the calculated conductance in the strongly spin-split case for $\varepsilon$ greater than, equal to, and less than zero.
For $\varepsilon = 0$, the Coulomb blockade is 1$e$-periodic---the onsets of the lowest cotunneling steps in both odd and even valleys align at $v=\Delta$. 
For $\varepsilon$ away from zero, the spectrum is even-odd periodic with the inelastic onset at $v=\Delta-\varepsilon$ in the odd and $v=\Delta+\varepsilon$ in the even valleys. The sign of $\varepsilon$ determines the relative size of odd and even Coulomb diamonds.

\onecolumngrid
\twocolumngrid
\newpage

\begin{figure*}[h!]
\includegraphics[width=\linewidth]{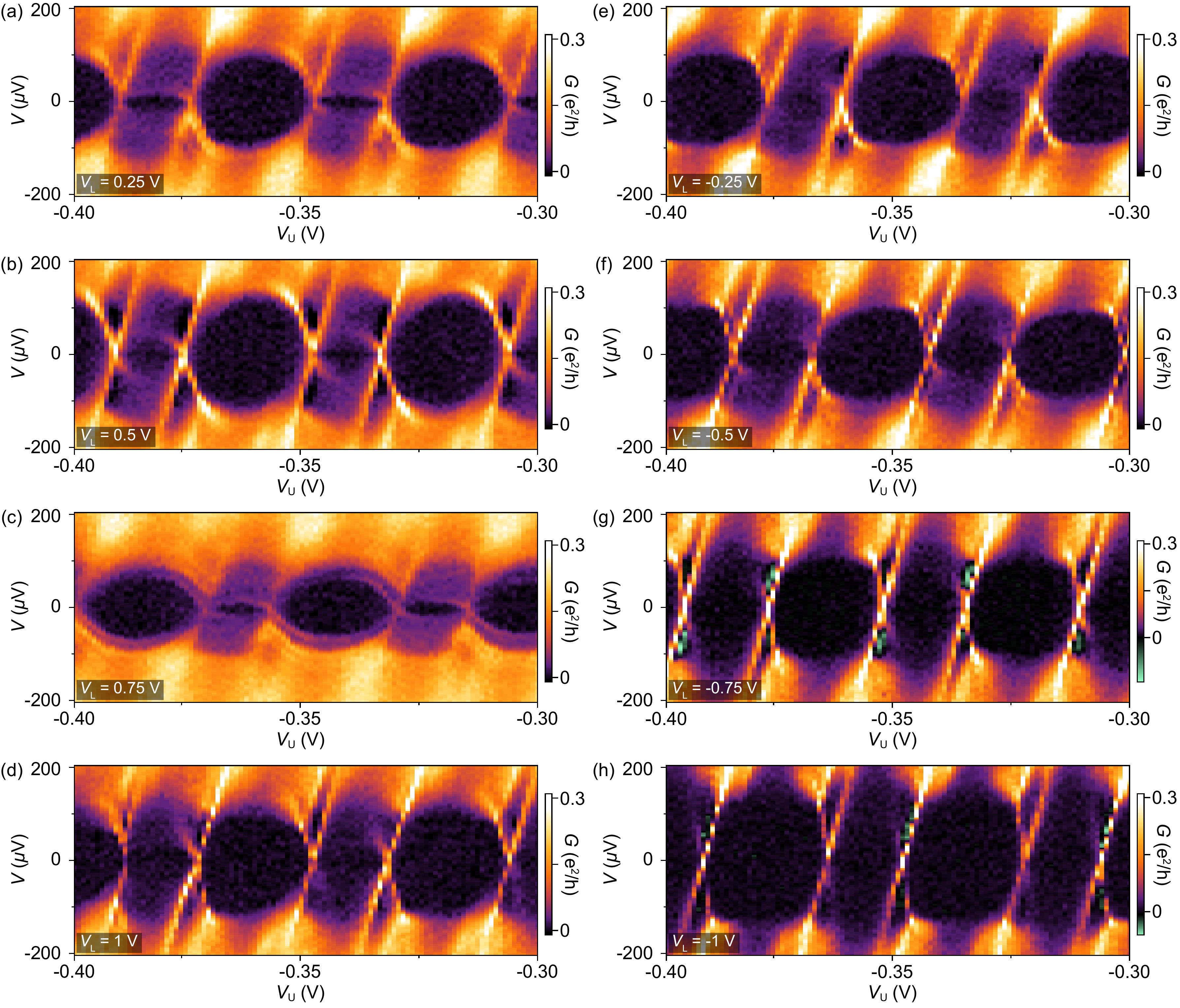}
\caption{\label{fig:S1} 
Differential conductance, $G$, measured for the 400~nm island on wire 1 at various lower-gate voltages, $V_{\rm L}$, as a function of bias, $V$, and upper-gate voltage, $V_{\rm U}$.
All spectra display even-odd periodic Coulomb blockade and a step in conductance around $V = 120~\mu$V.
Around zero and for positive $V_{\rm L}$ settings, additional steps in conductance at valley-dependent $V$ can be seen.
For more negative $V_{\rm L}$ these steps fade out, and negative differential conductance becomes apparent.
}
\end{figure*}

\begin{figure*}[h!]
\includegraphics[width=\linewidth]{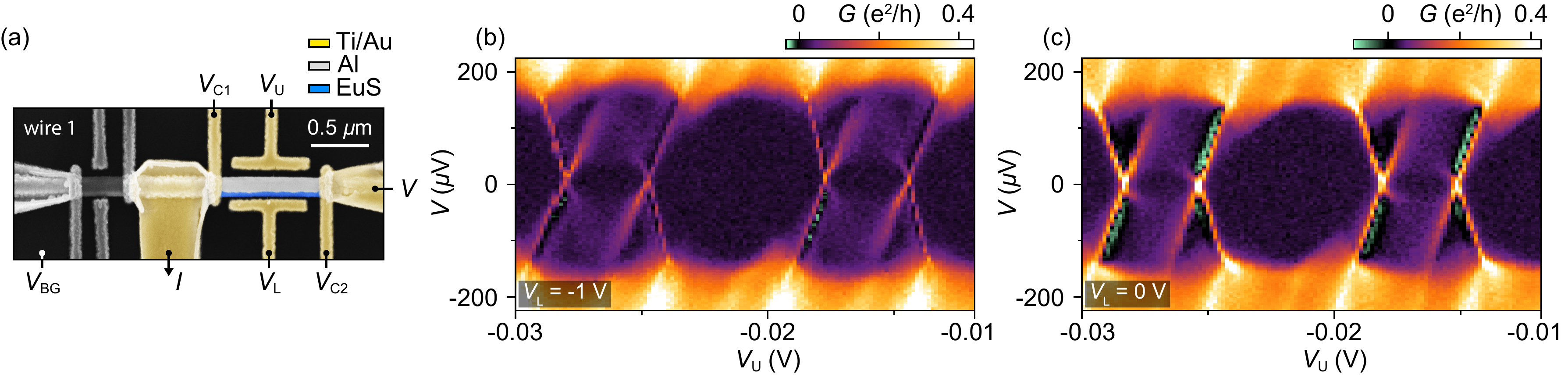}
\caption{\label{fig:S2}
(a) Scanning electron micrograph of wire~1 with color-enhanced setup for the longer, 800~nm island.
(b) Differential conductance, $G$, as a function of bias voltage, $V$, and upper-gate voltage, $V_{\rm U}$, shows even-odd alternation with the conductance-step onsets alternating between $V = 30$ and $140~\mu$V, as well as around $V = 170~\mu$V in both valleys.
The data were taken at $V_{\rm L} = -1$~V.
(c) Similar to (b) but taken at $V_{\rm L} = 0$ with the lower energy conductance-step onsets alternating between roughly $V = 20$ and $160~\mu$V from diamond to diamond and the higher energy onset around $V = 160~\mu$V in both valleys.
}
\end{figure*}

\begin{figure*}[h!]
\includegraphics[width=\linewidth]{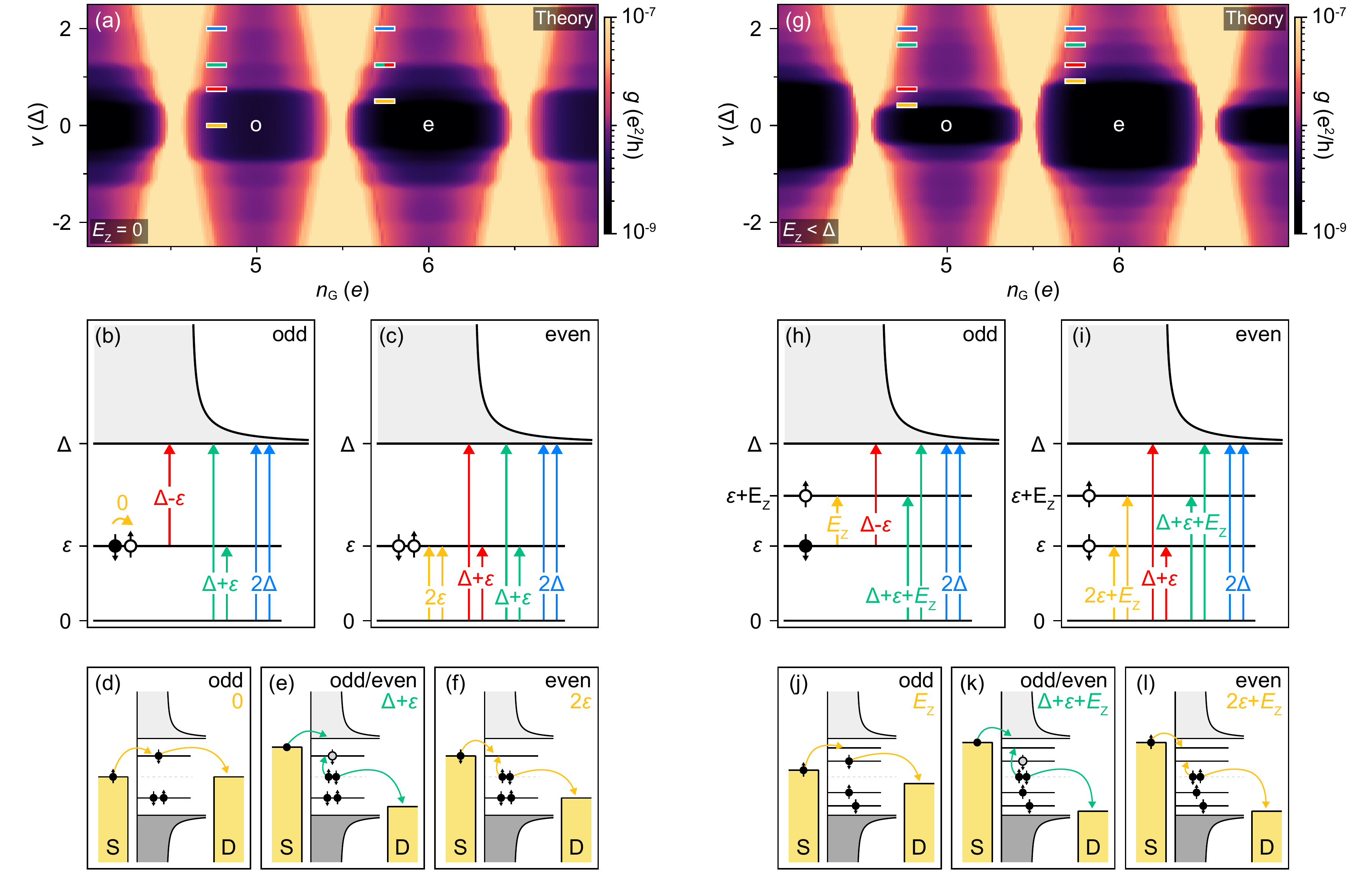}
\caption{\label{fig:S3}
(a) Calculated differential conductance, $g$, as a function bias, $v$, and offset charge, $n_{\rm G}$, for the model of a Coulomb island with a superconducting gap $\Delta$, charging energy $E_{\rm C} = 5\Delta$, a spin-degenerate subgap state $\varepsilon = \Delta/4$ with the Bogoliubov components $u_0=v_0=1/\sqrt{2}$, and Zeeman energy $E_{\rm Z} = 0$.
The odd valley displays a relatively higher background conductance down to $v=0$. 
Steps in $g$ are visible at $v = \Delta-\varepsilon$ in odd, and $v=2\varepsilon$ in even Coulomb valleys as well as at $v=\Delta+\varepsilon$ and $2\Delta$ in both Coulomb valleys.
The calculations were done for finite temperature $k_{\rm B}\,T=0.03\,\Delta$.
(b) and (c) Schematic representation of superconducting density of states for odd (b) and even (c) valleys in (a) with possible inelastic cotunneling excitations indicated by arrows.
(d) In the odd ground state, the quasiparticle in the bound state can exit the island, while another quasiparticle with an opposite spin from a lead tunnels back into the bound state. The process does not require energy as at $E_{\rm Z} = 0$ the subgap state is spin degenerate.
(e) In both occupancies, a Cooper pair can be broken-up with the spin-down quasiparticle leaving the island and the spin-up quasiparticle being excited into the spin-degenerate bound state, while another quasiparticle from a lead tunnels into the continuum. The total energy cost of the process is $\Delta+\varepsilon$.
(f) Similar to (e) but the quasiparticle entering from the lead is spin-down and tunnels into the bound state.
(g)-(l) Similar to (a)-(f) but for a weakly spin-split subgap state, offsetting the processes marked with yellow and green by $E_{\rm Z}$, while the processes marked with red and blue remain unaltered. The spectrum in (g) is calculated using $E_Z=0.42\,\Delta$, while the other parameters are the same parameters as in (a).
}
\end{figure*}

\begin{figure*}[h!]
\includegraphics[width=\linewidth]{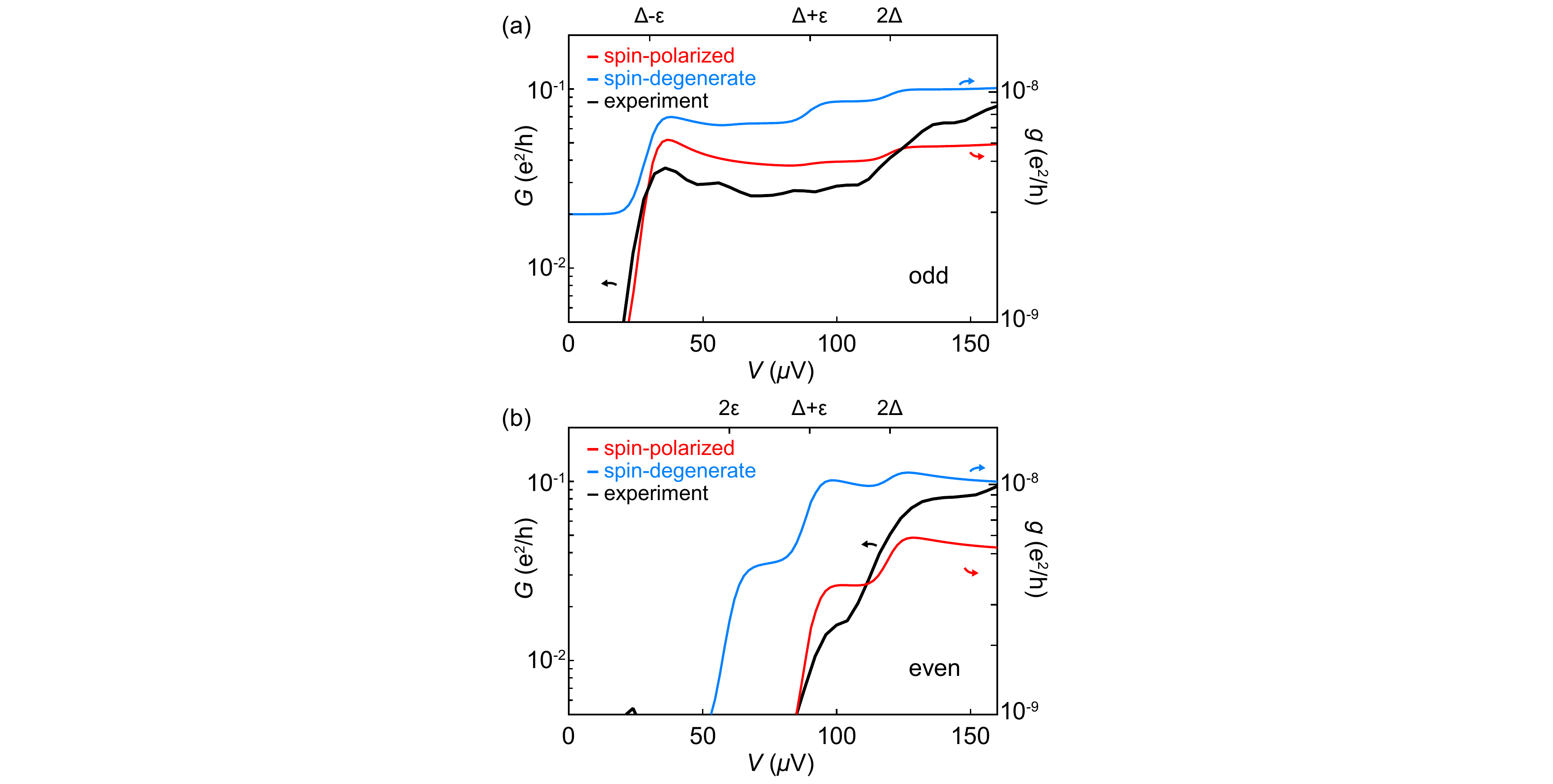}
\caption{\label{fig:S4} 
(a) Black curve, left axis: Differential conductance, $G$, line-cut taken from the main-text Fig.~1(c) at $V_{\rm U} = -0.474$~V, in the middle of a smaller Coulomb valley. $G$ increases in a step-like manner at $V=30~\mu$eV and then again around $120~\mu$eV.
Red curve, right axis: Calculated conductance, $g$, through a hybrid island with a superconducting gap $\Delta = 60~\mu$V, spin-polarized subgap state at $\varepsilon = 30~\mu$V, and an odd offset charge, $n_{\rm G} = 5$e, showing two cotunneling steps at $\Delta-\varepsilon$ and $2\Delta$.
Blue curve, right axis: Same as the red curve, but for a spin-degenerate bound state, showing a finite background conductance starting at $V=0$ as well as three cotunneling steps at $\Delta-\varepsilon$, $\Delta+\varepsilon$, and $2\Delta$.
(b) Same as (a) but the experimental data were taken from the middle of a larger Coulomb valley at $V_{\rm U} = -0.409$~V, showing two step-like increases at $V=90$ and $120~\mu$V; the calculations were done for an even offset charge, $n_{\rm G} = 6$e, showing two cotunneling steps at $\Delta+\varepsilon$ and $2\Delta$ for the spin-polarized case, and an additional third step at $2\varepsilon$ in the spin-degenerate case.
}
\end{figure*}

\begin{figure*}[h!]
\includegraphics[width=\linewidth]{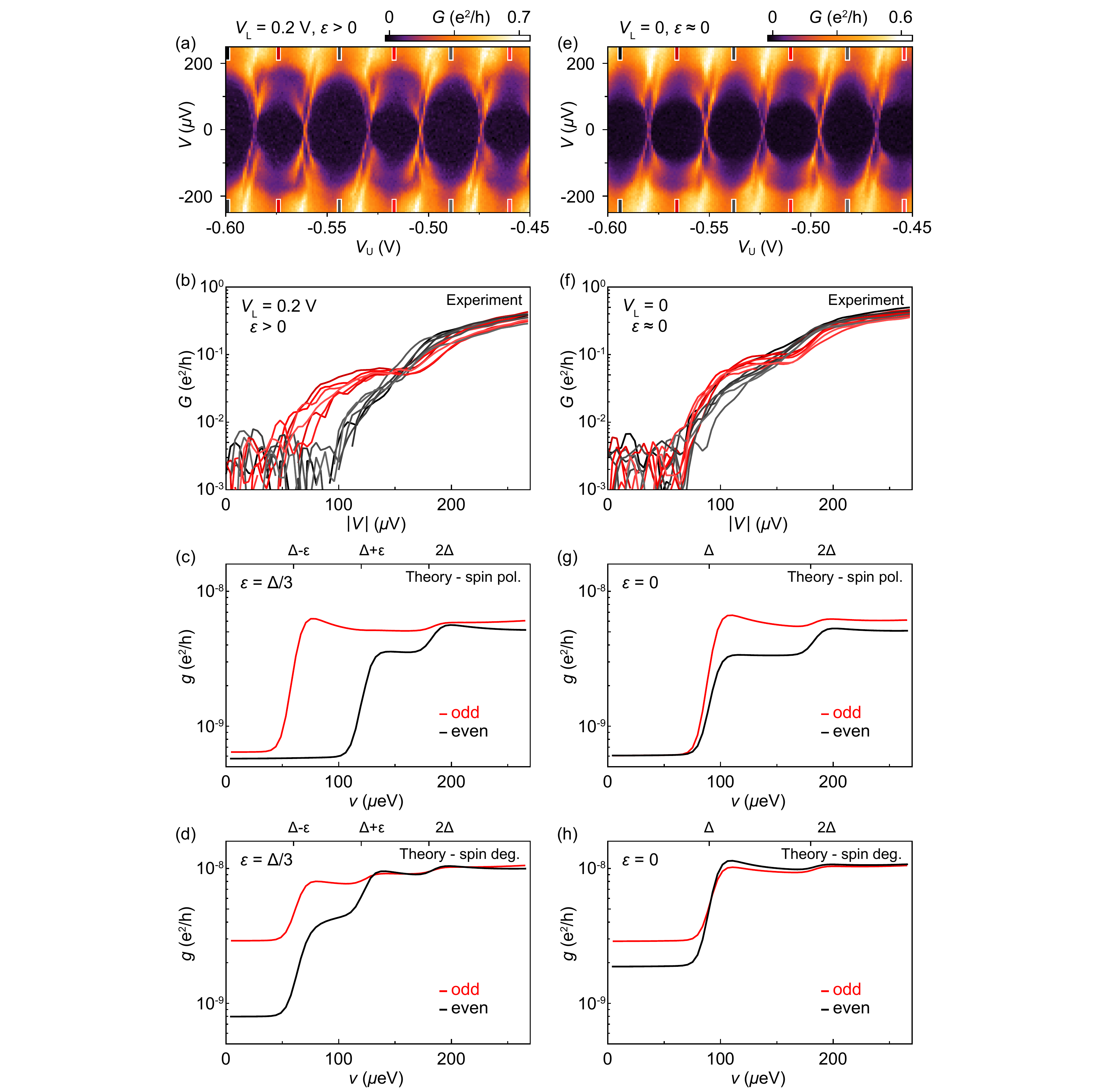}
\caption{\label{fig:S5} 
(a) Same data as in the main-text Fig.~3(a): Differential conductance, $G$, as a function of source-drain bias, $V$, and upper gate voltage,\ $V_{\rm U}$, for the 400 nm island on wire 2. The data were taken at a fixed lower-gate voltage $V_{\rm L} = 0.2$~V. A clear even-odd Coulomb blockade pattern displays two cotunneling steps in both valleys: around 60 and 180~$\mu$eV in the smaller, and 120 and 180~$\mu$eV in the larger valley.
(b)~Conductance line-cuts taken from (a) at $V_{\rm U}$ values marked by the ticks, plotted against the absolute value of $V$. Red~(black) curves are from the smaller (larger) Coulomb diamonds.
(c) Calculated conductance, $g$, through a hybrid island with a superconducting gap $\Delta = 90~\mu$V and spin-polarized subgap state at $\varepsilon = 30~\mu$V, for even (black) and odd (red) offset charges. Both curves display two cotunneling steps at $\Delta+\varepsilon$ and $2\Delta$ for even occupancy, and $\Delta-\varepsilon$ and $2\Delta$ for odd occupancy.
(d) Similar to (c) but for a spin-degenerate subgap state. Odd curve displays a relatively larger background conductance. Both even and odd curves display three cotunneling steps at $\Delta-\varepsilon$, $\Delta+\varepsilon$, and $2\Delta$.
(e)-(h)~Similar to (a)-(d) but for $V_{\rm L} = 0$ and $\varepsilon\approx0$. The Coulomb blockade is 1$e$ periodic, with all valleys displaying two cotunneling steps around 90 and 180~$\mu$eV.
Independent of lower-gate voltage, the measured data display only two cotunneling steps in all Coulomb valleys, with comparable background conductances, suggesting transport trough spin-polarized subgap states.
}
\end{figure*}

\begin{figure*}[h!]
\includegraphics[width=\linewidth]{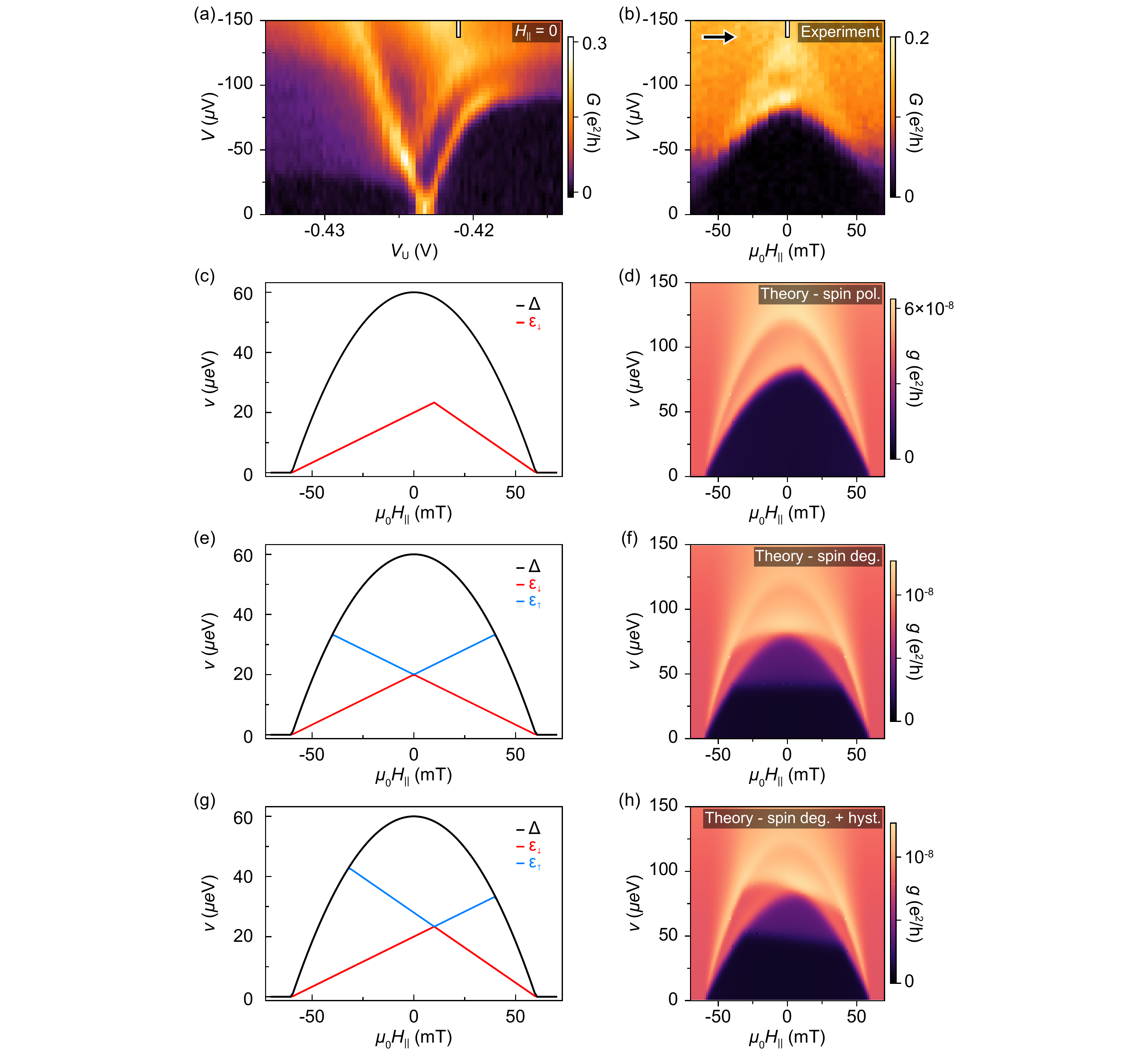}
\caption{\label{fig:S6} 
(a) Differencial conductance, $G$, as a function of voltage bias, $V$, and upper-gate voltage, $V_{\rm U}$, for the 400 nm island on wire 1 at zero applied magnetic field and the same gate configuration as in the main-text Fig.~1(c).
(b) $G$ as a function of $V$ and external magnetic field, $\mu_0 H_\parallel$, applied parallel to the wire axis, measured at $V_{\rm U} = -0.421$~V. Sweep direction indicated by arrow. The onsets of the two conductance steps decrease in $V$ for $H_\parallel$ away from zero, but do not split. A small hysteresis of $\sim 10$~mT can be seen.
(c) Superconducting gap, $\Delta$, and the spin-up branch of a subgap state, $\varepsilon_\uparrow$, values as a function of an $H_\parallel$, that were used as an input for spin-polarized model. In this case the subgap state is assumed to display 10~mT hysteresis.
(d) Calculated conductance, $g$, through a hybrid island with a spin-polarized subgap state and an even parity, as a function of $H_\parallel$, taking $\Delta$ and $\varepsilon_\uparrow$ from (c). The spectrum displays only two cotunneling features throughout the superconducting range.
(e) and (g) Similar to (c) and (d) but assuming a subgap state that is spin-degenerate at $H_\parallel = 0$. The spectrum in (d) exhibits three cotunneling features that split in field.
(g) and (h) Similar to (e) and (g) but with 10~mT hysteresis, showing a complicated pattern of up to four cotunneling features.
}
\end{figure*}

\begin{figure*}[h!]
\includegraphics[width=\linewidth]{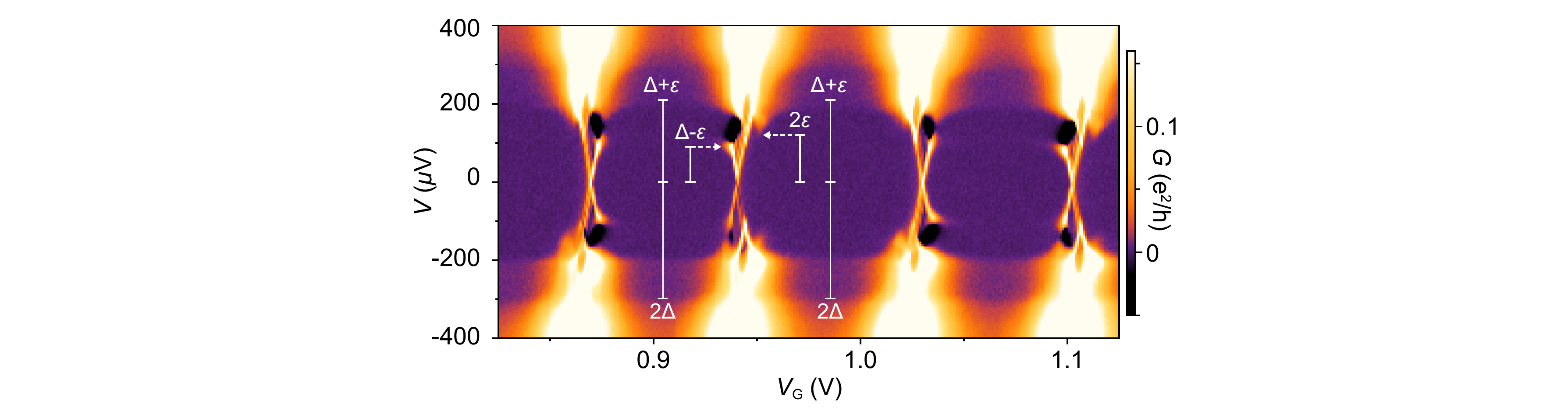}
\caption{\label{fig:S7} 
(a) Differential conductance, $G$, for a 310 nm InAs/Al hybrid island without EuS as a function of gate voltage, $V_{\rm G}$, and source–drain voltage, $V$, at zero external magnetic field.
Two clear steps in conductance can be seen for both charge occupancies around $V=210$ and $300~\mu$eV. We associate these features with excitations at $\Delta+\varepsilon$ and $2\Delta$, respectively, yielding $\Delta = 150~\mu$eV and $\varepsilon = 60~\mu$eV. Two additional, less pronounced features at the edges of the Coulomb diamonds (see dashed arrows) alternate between $V = 90$ and $120~\mu$eV and can be associated with $\Delta-\varepsilon$ and $2\varepsilon$, respectively. The data are replotted from Ref.~\cite{Higginbotham2015}.
}
\end{figure*}

\begin{figure*}[h!]
\includegraphics[width=\linewidth]{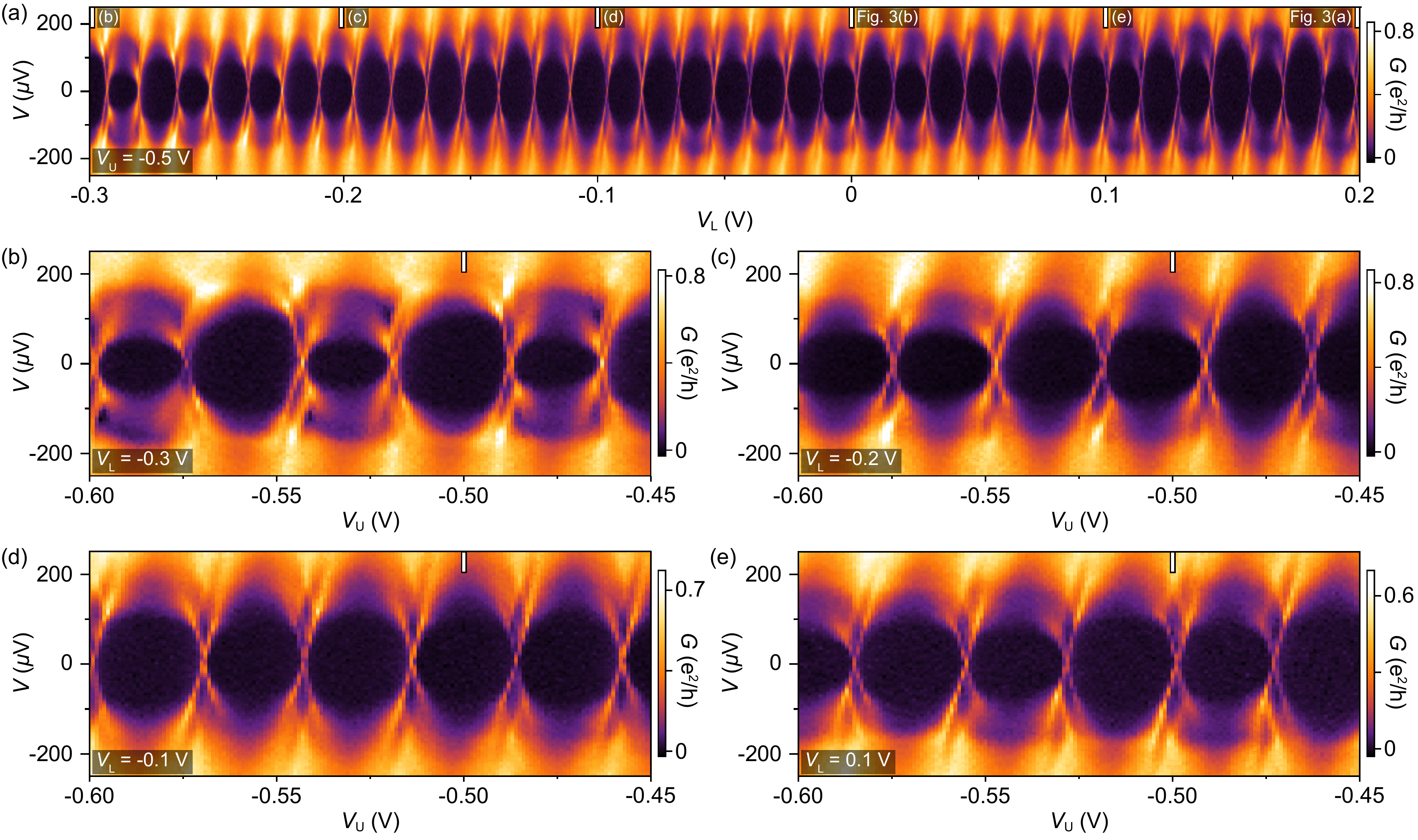}
\caption{\label{fig:S8} 
(a) Differential conductance, $G$, measured for the 400~nm island on wire 2 as a function of bias, $V$, and lower-gate voltage, $V_{\rm L}$, which is positioned on the side of the wire with the EuS shell [see the main-text Fig.~1(a)]. This gate tunes carrier density in the semiconductor as well as the island occupancy. The data span 36 Coulomb diamonds, taken at fixed $V_{\rm U} = -0.5$~V.
(b) $G$ dependence on $V_{\rm U}$ taken at $V_{\rm L} =$ -0.3~V.
(c)-(e) Same as (b) but measured at $V_{\rm L} =$ -0.2~V (c), -0.1~V (d), and 0.1~V (e). Similar data taken at $V_{\rm L} = 0$ and 0.2~V are shown in the main-text Fig.~3.
}
\end{figure*}

\begin{figure*}[h!]
\includegraphics[width=\linewidth]{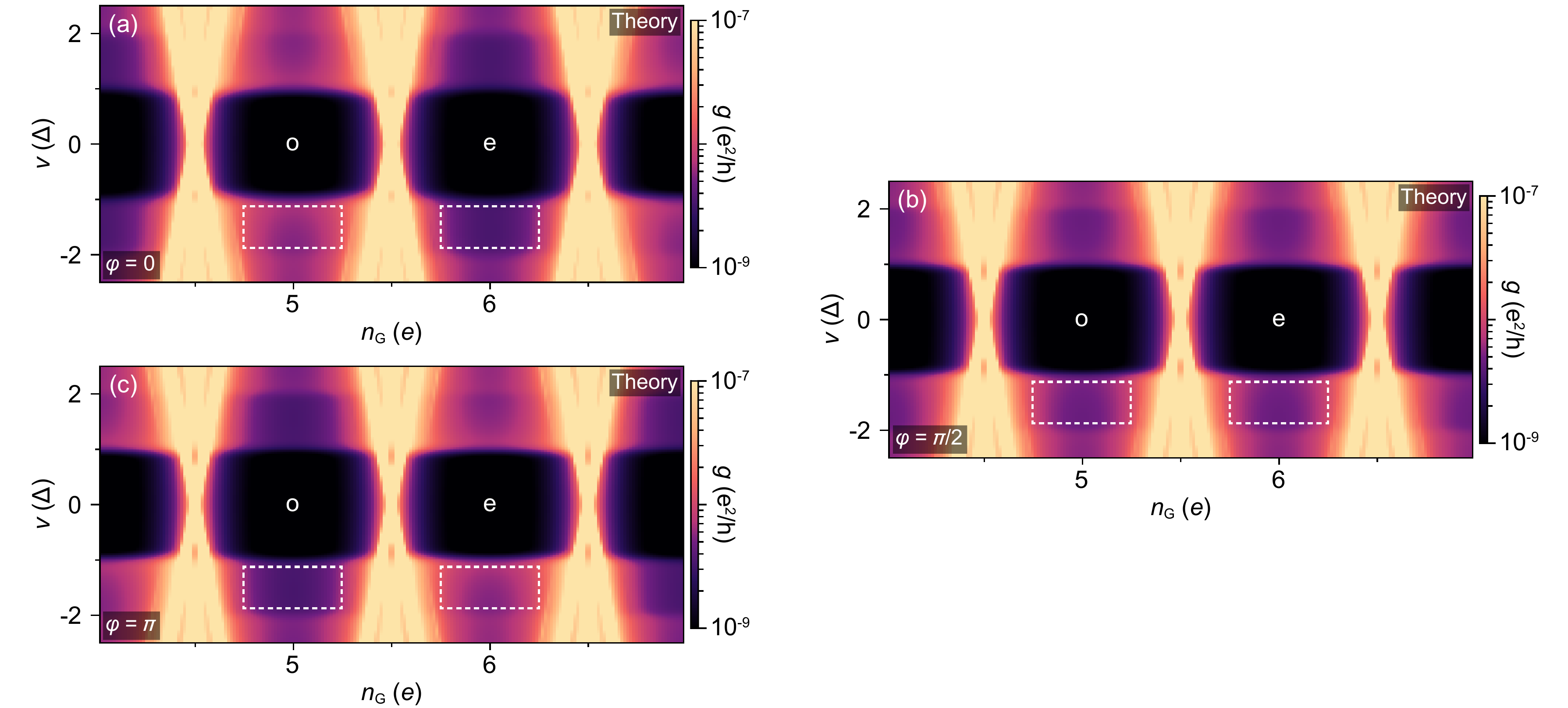}
\caption{\label{fig:S9} 
Calculated differential conductance, $g$, as a function bias, $v$, and offset charge, $n_{\rm G}$, for a superconducting island with bound state at $\varepsilon=0$ with the relative phase between the Bogoliubov components $\varphi = 0$ (a), $\pi/2$ (b), and $\pi$ (c), where $\varphi$ is given by $u_0=\exp(i\varphi)v_0$. The other parameters are the same as in the main-text Fig.~2(a). Depending on the value of $\varphi$, the amplitude of the conductance step at $v=\Delta$ in the odd valley can be (a) higher than, (b) equal to, or (c) lower than the step in the even valley.
}
\end{figure*}

\begin{figure*}[h!]
\includegraphics[width=\linewidth]{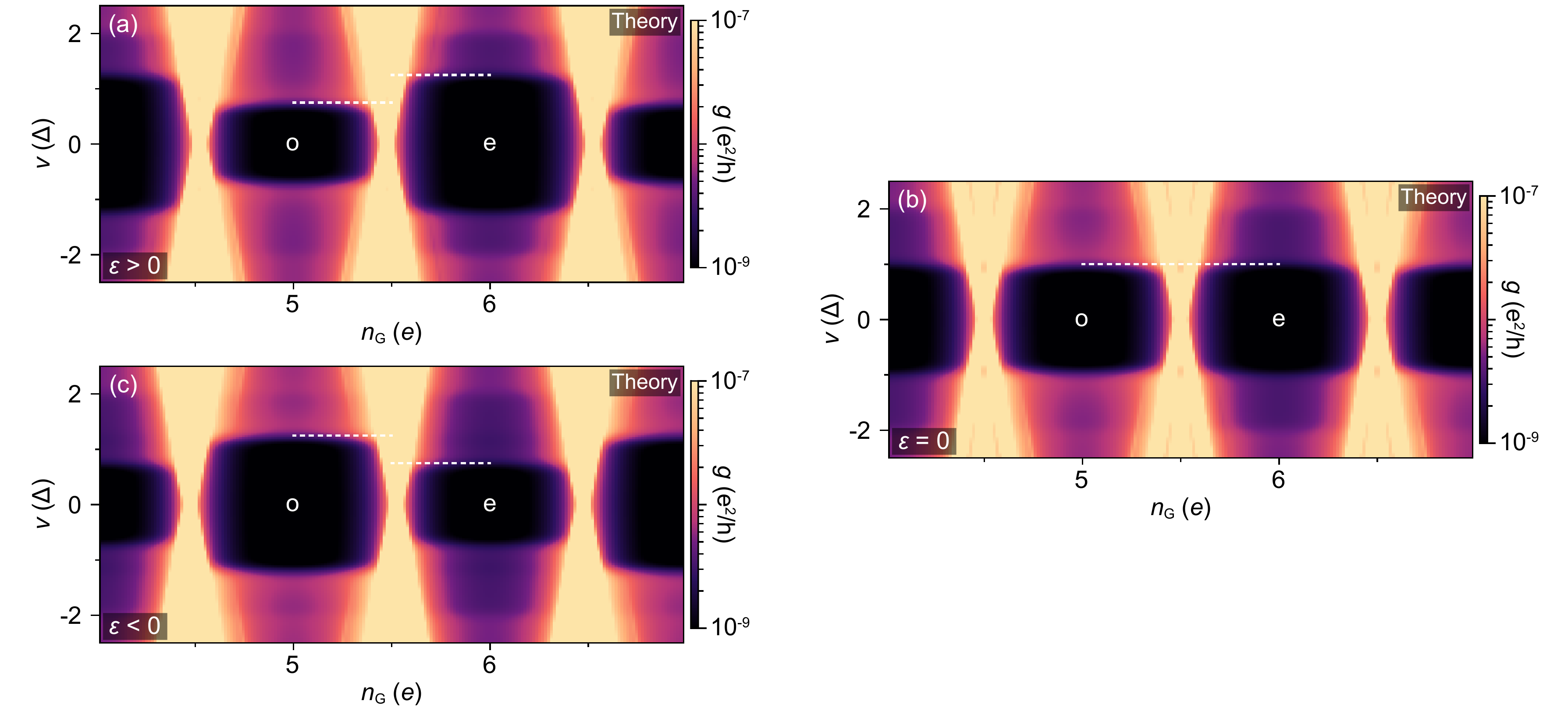}
\caption{\label{fig:S10} 
Calculated differential conductance, $g$, as a function of bias, $v$, and offset charge, $n_{\rm G}$, for a superconducting island with a bound state at $\varepsilon=\Delta/4$ (a), $0$ (b), and $-\Delta/4$ (c). The other parameters are the same as in the main-text Fig.~2(a). Depending on the value of $\varepsilon$, the lowest inelastic onset in the odd valley can be at (a) lower, (b) same, or (c) higher value of $v$ compared to the lowest cotunneling onset in the even valley. 
}
\end{figure*}

\begin{figure*}[h!]
\includegraphics[width=\linewidth]{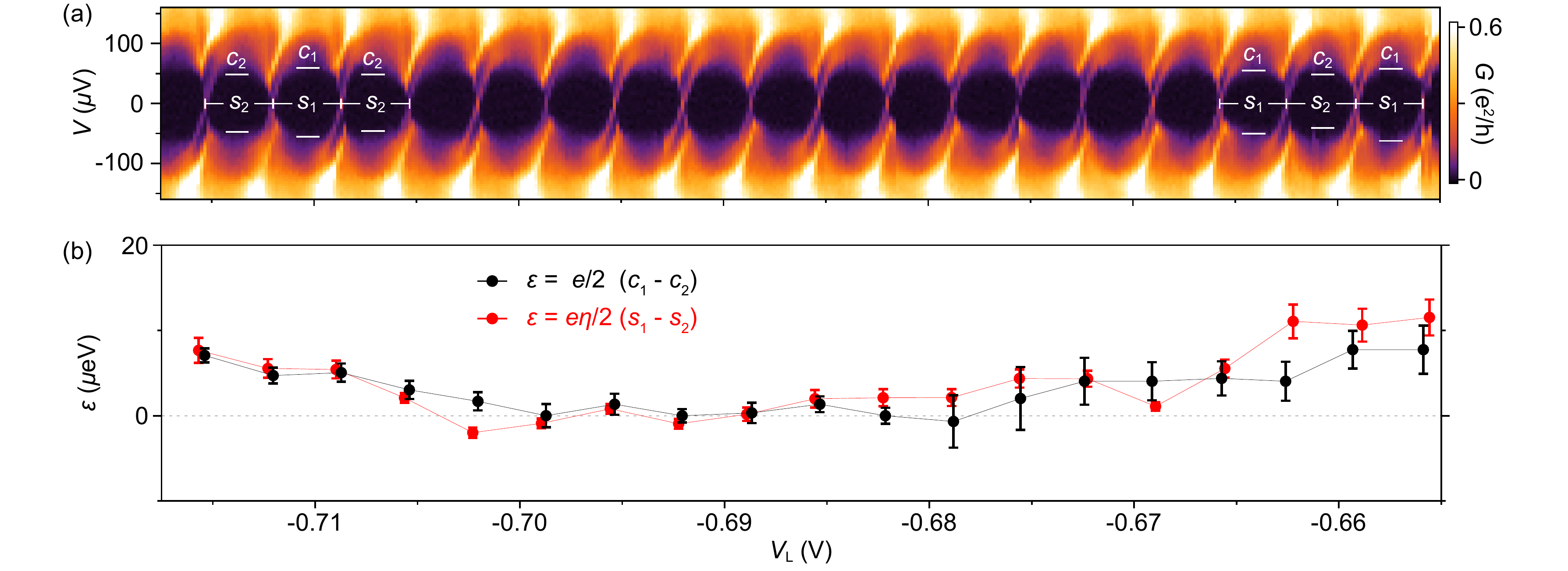}
\caption{\label{fig:S11} 
(a) Differential conductance, $G$, for the 800 nm island on wire 2 as a function of bias, $V$, over an extended range of  lower-gate voltage, $V_{\rm L}$. The Coulomb blockade pattern evolves from even-odd at $V_{\rm L} = -0.7$~V, through 1$e$ around $V_{\rm L} = -0.69$~V, to even-odd periodicity again at $V_{\rm L} = -0.67$~V, visible in both inelastic cotunneling onsets $c_{\rm i}$ and peak spacings $s_{\rm i}$, where $i = 1$ and 2 denote Coulomb valleys with the same charge occupancies.
$i=1$ diamonds are larger than the $i=2$ diamonds on both ends of the measured gate voltage range.
The data were taken at $V_{\rm U} = -1.43$~V.
(b) $\varepsilon$ deduced from the data shown in (a) using the inelastic cotunneling onsets (black) and the peak-spacing difference (red).
$\varepsilon$ decreases from roughly $10~\mu$eV to 0, but then increases again to $10~\mu$eV as the gate voltage is increased.
The black error bars represent standard errors from the $c_{\rm i}$ measurement at the positive and negative $V$, whereas the red error bars were estimated by propagation of error from Lorentzian peak fitting and lever arm, $\eta$, measurement. 
}
\end{figure*}

\begin{figure*}[h!]
\includegraphics[width=\linewidth]{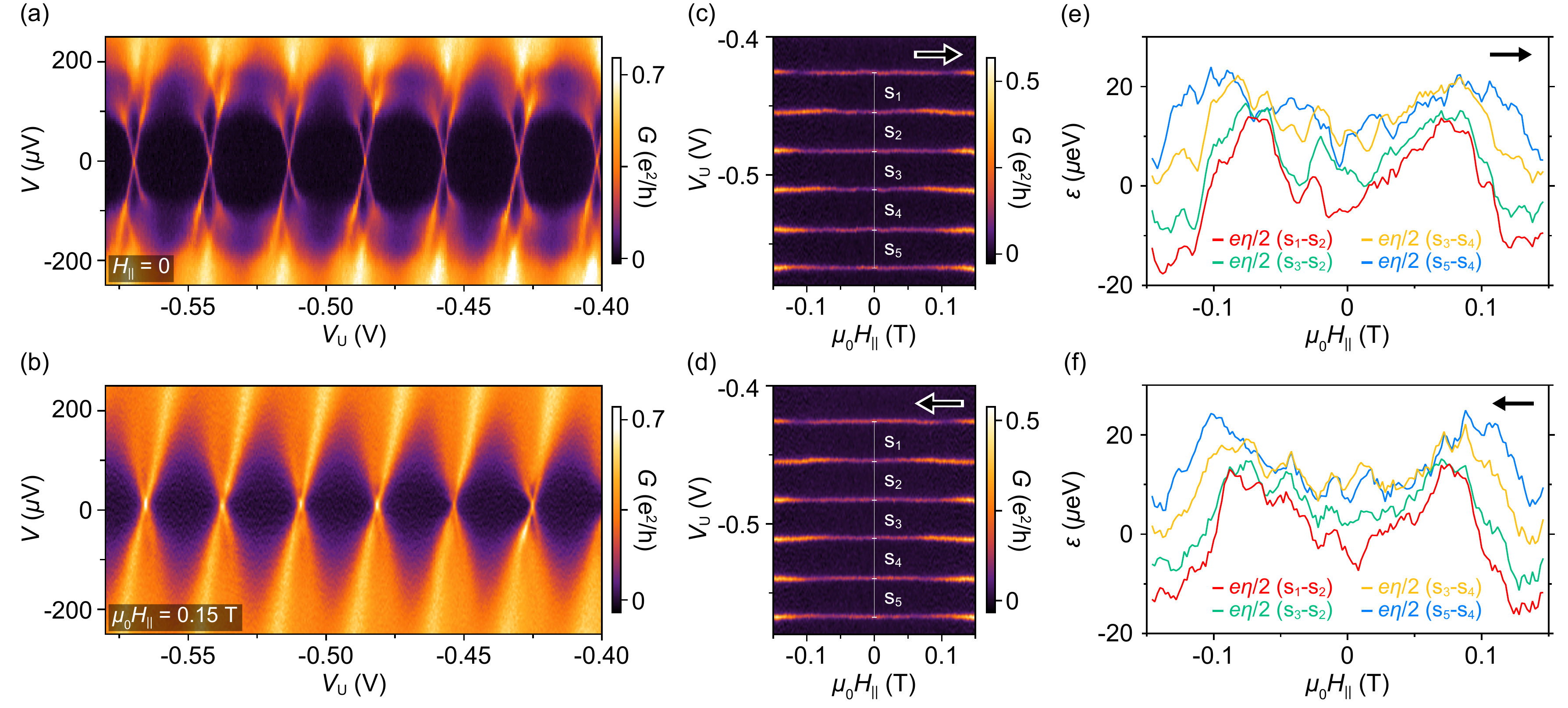}
\caption{\label{fig:S12} 
(a) Differential conductance, $G$, as a function of bias, $V$, and upper-gate voltage, $V_{\rm U}$, for the 400~nm island on wire 2, at zero applied magnetic field and lower-gate voltage $V_{\rm L} = 0$. Coulomb blockade is nearly 1$e$ periodic with finite bias conductance steps at $V = 90$ and $180~\mu$eV in each diamond as well as discrete peaks at the degeneracy points.
(b) Similar to (a) taken at finite external magnetic field $\mu_0 H_\parallel = 150$~mT applied parallel to the wire axis. Coulomb blockade shows finite, featureless conductance outside of the diamonds and no sign of excited states.
(c) Zero-bias $G$ as a function of $V_{\rm U}$ and $\mu_0 H_\parallel$. Sweep direction indicated by arrow. Both the peak amplitude and position in $V_{\rm U}$ changes nonmonotonically with $H_\parallel$.
(d) Same as (c) with sweep direction from positive to negative field.
(e) $\varepsilon$ deduced from the difference of two consecutive peak spacings indicated in (c) allowing to track the subgap state evolution with both $H_\parallel$ and $V_{\rm U}$.
(f) Same as (e) but for the data from (d).
}
\end{figure*}

\begin{figure*}[h!]
\includegraphics[width=\linewidth]{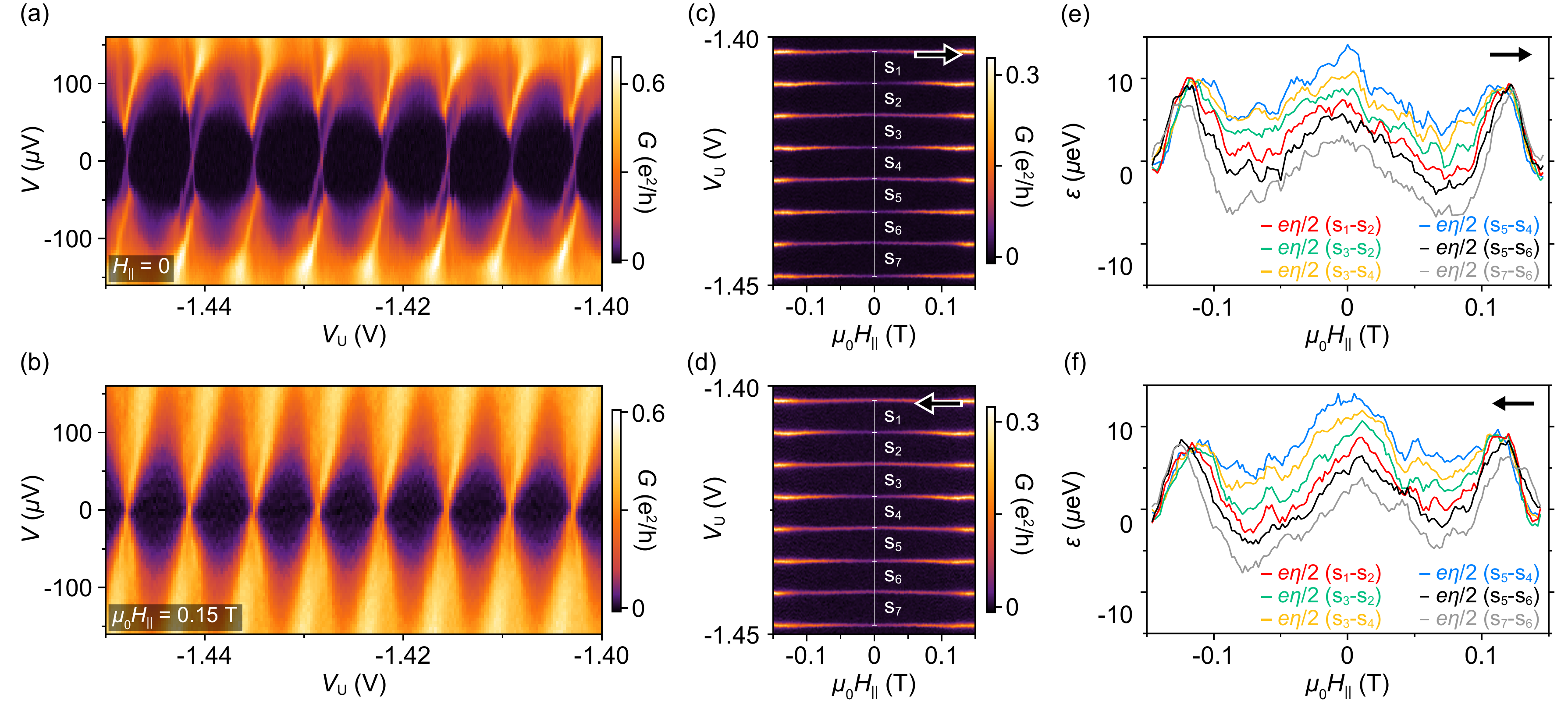}
\caption{\label{fig:S13} 
(a) Differential conductance, $G$, as a function of bias, $V$, and upper-gate voltage, $V_{\rm U}$, for the 800~nm island on wire 2, at zero applied magnetic field and lower-gate voltage $V_{\rm L} = -0.7$~V. Coulomb blockade is nearly 1$e$ periodic with two inelastic cotunneling onsets at $V = 60$ and $120~\mu$eV in each diamond as well as discrete peaks at the degeneracy points.
(b) Similar to (a) taken at $\mu_0 H_\parallel = 150$~mT. Coulomb blockade shows finite, featureless conductance outside of the diamonds and no sign of excited states.
(c) Zero-bias $G$ as a function of $V_{\rm U}$ and $\mu_0 H_\parallel$. Sweep direction indicated by arrow. Both the peak amplitude and position in $V_{\rm U}$ changes nonmonotonically with $H_\parallel$.
(d) Same as (c) with sweep direction from positive to negative field.
(e) $\varepsilon$ deduced from the difference of two consecutive peak spacings indicated in (c) allowing to track the subgap state evolution with both $H_\parallel$ and $V_{\rm U}$.
(f) Same as (e) but for the data from (d).
}
\end{figure*}

\begin{figure*}[h!]
\includegraphics[width=\linewidth]{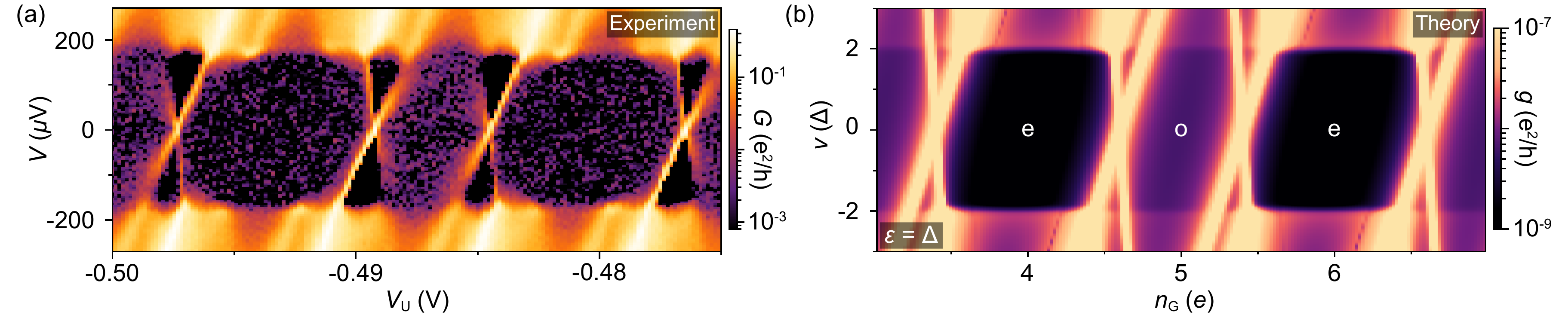}
\caption{\label{fig:S14} 
(a) Differential conductivity, $G$, as a function of bias, $V$, and upper-gate voltage, $V_{\rm U}$, for the 800~nm island on wire 2, at lower-gate voltage $V_{\rm L} = 0$~V. The spectrum shows an even-odd periodic Coulomb blockade with the conductance suppressed below the noise floor in the larger valleys and small, but finite conductance throughout the smaller valleys.
No further features are observed up to roughly $V=180~\mu$eV where $G$ increases in both Coulomb diamonds.
Note the logarithmic color scale.
(b) Calculated differential conductance, $g$, as a function of bias, $v$, and offset charge, $n_{\rm G}$, for a modeled superconducting Coulomb island with $\varepsilon=\Delta$, $E_{\rm Z}=0$, $E_{\rm C}=5\Delta$, and asymmetric voltage bias given by the left- and right-lead potential ratio $\mu_L/\mu_R=5$. The spectrum qualitatively agrees with the experimental data in (a).
}
\end{figure*}


\end{document}